\documentclass[onecolumn, draftclsnofoot, 12pt]{IEEEtran} 
\usepackage{preamble}
\usepackage{cite}

\begin{document}
\title{Joint Communication and Parameter Estimation\\in MIMO Channels}
\author{
    Gökhan~Yılmaz, 
    Franz~Lampel, 
    Hamdi~Joudeh 
    and~Giuseppe~Caire
    \thanks{
        Gökhan Yılmaz and Hamdi Joudeh are with the Department of Electrical Engineering, Eindhoven University of Technology, 5600 MB Eindhoven, the Netherlands (e-mail: g.yilmaz@tue.nl; h.joudeh@tue.nl). 
    }%
    \thanks{
        Franz Lampel was with the Department of Electrical Engineering, Eindhoven University of Technology, 5600 MB Eindhoven, the Netherlands. He is now with Synopsys, Inc., High Tech Campus 41, 5656 AE Eindhoven, the Netherlands (e-mail: lampel@synopsys.com).
    }
    \thanks{
        Giuseppe Caire is with the Faculty of Electrical Engineering and Computer Science, Technical University of Berlin, 10623 Berlin, Germany (e-mail: caire@tu-berlin.de).
    }%
    \thanks{
        This work was supported by the European Research Council (ERC) under Grant 101116550.
        The work of Giuseppe Caire was supported by the Gottfried Wilhelm Leibniz-Preis 2021 of the German Science Foundation (DFG).
    }%
}
\maketitle
\begin{abstract}
We study a joint communication and sensing setting comprising a transmitter, a receiver, and a sensor, all equipped with multiple antennas.
The transmitter sends an encoded signal over the channel with the dual purpose of communicating an information message to the receiver, and enabling the sensor to estimate a target parameter vector by generating back-scattered signals.
We assume that the transmitter and sensor are co-located, or fully connected, giving the latter access to the transmitted signal.
The target parameter vector is randomly drawn from a continuous distribution, yet remains fixed throughout the transmission block.
We establish the fundamental performance trade-off between the communication and sensing tasks, captured in terms of a capacity-MSE function. 
In doing so, we identify optimal coding schemes for this multi-antenna joint communication and sensing setting.
Moreover, we particularize our result to two practically-inspired scenarios where we showcase optimal schemes and trade-offs.
\end{abstract}
\section{Introduction}
\label{sec:intro}
Joint communication and sensing (JCAS), also known as integrated sensing and communication (ISAC), is expected to be a key feature of next-generation wireless technology.
This emerging paradigm aims at designing efficient wireless systems with transceivers that can utilize the same physical resources (e.g., hardware, spectrum, and power) to carry out both communication and sensing tasks \cite{fanliu2020,fliu2018, fliu2018_2, xliu2020, chen2021}.
Envisaged use cases include cellular networks, where base stations will have the ability to communicate with active devices and simultaneously detect and track passive targets from back-scattered signals \cite{Wild2021}; and automotive, where vehicles will be able to sense their surroundings to identify and track road obstacles, while communicating with other vehicles and the infrastructure \cite{Sturm2011,Ma2020}.
In both of these use cases, and several others, the underlying JCAS physical channel will very likely be a multiple-input and multiple-output (MIMO) channel, due to the employment of multi-antenna terminals, multiple subcarriers, or both.

In recent years, a number of studies have emerged investigating various aspects of MIMO in JCAS; see, e.g., \cite{fliu2018,fliu2018_2,xliu2020,chen2021,fliu2022,hua2024,ren2024,xu2024,xiong2023,joudeh2023}.
With the exception of \cite{xiong2023,joudeh2023}, the vast majority of studies on MIMO JCAS focus mainly on signal processing and optimization aspects.
In particular, a transmission scheme with a predetermined signal structure, often based on linear precoding and Gaussian signaling, is adopted; reducing the problem to the design of signal parameters (e.g., covariance matrices) to achieve different performance trade-offs.
While these works have provided useful insights into the potential benefits of co-designed MIMO JCAS systems, without an information-theoretic treatment involving coding theorems and converse bounds, the optimality of these predetermined signal structures and schemes remains unclear. Moreover, we still lack performance benchmarks that indicate how much more can be gained.
\subsubsection*{Goal and considered setting} 
Our primary goal in this work is to shed light on the fundamental performance limits of MIMO JCAS systems by taking an information-theoretic approach. To this end, we focus on an elemental setting comprising a transmitter, a receiver, and a sensor, all equipped with multiple antennas (or any vector dimension, e.g., subcarriers). 
The transmitter and sensor are co-located or strongly connected, enabling monostatic sensing, and henceforth can be thought of as components of the same cellular base station (BS). The receiver, on the other hand, belongs to an active user equipment served by the BS.
The transmitter sends an encoded signal with the dual purpose of communicating a message to the receiver and, at the same time, probing the surrounding environment and generating a back-scatter signal to enable sensing.
We refer to this setting as a \emph{MIMO JCAS channel}.

Information-theoretic formulations and performance limits for MIMO \emph{communication} channels are well established and extensively studied \cite{Telatar1999, Foschini1998, Goldsmith2003}.
To treat MIMO JCAS channels with the same degree of rigor, it is crucial to concretely formulate the \emph{sensing} aspect of the problem.
In the current work, we focus on the case where the sensing task involves estimating a \emph{continuous} parameter vector of the back-scatter channel.
The parameter vector is drawn at random according to a known prior distribution, yet remains \emph{fixed} throughout the transmission block.
This \emph{abstraction} is meant to model scenarios where parameters of interest change at a much slower timescale compared to channel uses (i.e., symbols).
Practical examples include direction-of-arrival (DoA), range (delay), and velocity (Doppler) estimation, all scenarios in which only slight changes in parameters occur over the span of a typical transmission block (see Remark \ref{remark:asymptotics}).
\subsubsection*{Fixed vs. varying parameter} 
Our choice to adopt a fixed random parameter model for sensing stands in sharp contrast to a prevalent approach in the information-theoretic literature on JCAS, where the parameter of interest is assumed to vary in an i.i.d. fashion from one channel symbol to another.
The i.i.d. model was proposed by Kobayashi \emph{et al.} in \cite{kobayashi2018}, where the focus is on discrete memoryless and basic Gaussian channels; and has since then been widely adopted in several follow-up works \cite{kobayashi2019,ahmadipour2023,ahmadipour2024,gunlu2023,joudeh2024}.
In \cite{xiong2023}, Xiong \emph{et al.} study a MIMO JCAS channel that is very similar to ours, with the exception that their sensing model also follows the i.i.d. philosophy in \cite{kobayashi2018}.
In particular, they consider the estimation of a continuous parameter vector that varies across channel symbols (or sub-blocks of symbols) in an i.i.d. fashion. An asymptotically large number of symbols (or sub-blocks) is considered, resembling ergodic fading in wireless communication.
Moreover, to measure the sensing performance, a form of Bayesian Cramér-Rao bound (BCRB) \cite{Miller1978} is used in \cite{xiong2023} as a proxy for the mean square error (MSE) averaged over parameter states.

The BCRB is known to be generally not tight, even asymptotically, except in certain special cases \cite{vanTrees2013}; hence, the performances characterized in \cite{xiong2023}  are not exact in general.
Nevertheless, high signal-to-noise ratio (SNR) analysis reveals two distinct types of trade-offs in the MIMO JCAS channel model of
\cite{xiong2023}. 
The first is referred to as the \emph{deterministic–random trade-off (DRT)}, which arises when the family of Gaussian input distributions—optimal for communication alone—fails to achieve all Pareto-optimal performance trade-off points. In this case, ``less random"  input distributions are required to achieve trade-off points that favor sensing.
The second is referred to as the \emph{subspace trade-off (ST)}, which arises from the fact that the communication user and the sensing target often occupy distinct subspaces in the MIMO vector space. Hence, signals more aligned with the user (resp. target) favor communication (resp. sensing).

The i.i.d. model does not fully capture practical scenarios where the parameters of interest remain relatively constant within transmission blocks, as the ones highlighted earlier.
The idea to study fundamental performance limits in JCAS channels with a fixed parameter originated from Joudeh and Willems in \cite{joudeh2022}, where the focus is on the basic sensing task of target detection (i.e., a binary parameter).
Extensions later followed in  \cite{joudeh2023,wu2024,chang2023,seo2025b}, where the focus remained on discrete (mostly binary) parameters.
The extension to MIMO JCAS channels in \cite{joudeh2023} is particularly relevant, and can be seen as a detection counterpart to the estimation problem we consider in the present paper. 
In a recent contribution, Seo \cite{seo2025a} extended the fixed discrete parameter JCAS setting in \cite{joudeh2022,wu2024,chang2023} to the continuous parameter regime.
Our current paper follows in the footsteps of \cite{seo2025a} and further extends the approach therein in several directions, as discussed in detail below in light of our contributions.
\subsubsection*{Overview and contribution} 
We consider a MIMO JCAS channel where the sensing task is to estimate a random yet fixed parameter vector, as described earlier.
We adopt a Bayesian framework for parameter estimation, with an MSE-based risk function \cite{vanTrees2013,kay1993}.
For this setting, we establish the fundamental trade-off between communication and sensing efficiencies in the asymptotic regime of large block-length.
Communication efficiency is measured in terms of the rate of reliable communication, in bits per channel use; whose ultimate asymptotic limit is the celebrated Shannon capacity \cite{Cover}.
Sensing efficiency, on the other hand, is measured in terms of the MSE risk scaled by the block-length, which we refer to as the \emph{asymptotic MSE}.
This choice of scaling is commonplace in asymptotic statistics, as pointed out in \cite[Sec.~I]{Wu2011}, since the MSE risk in such settings decays at a rate proportional to one over the block-length.
Therefore, a larger asymptotic MSE implies a slower decay rate and vice versa.

In the main result of this paper, presented in Section \ref{sec:main}, we derive the exact optimal trade-off between the rate of reliable communication and the sensing asymptotic MSE, described in terms of a capacity-MSE function. The rate is characterized by a mutual information term, while the asymptotic MSE is characterized by a conditional variant of the Expected Cramér-Rao bound (ECRB) \cite{vanTrees2013}, adapted to our sensing setting. 
It is interesting to note that while the ECRB does not generally yield a valid lower bound at finite block-lengths, it is asymptotically tight \cite{vanTrees2013,aharon2024}. 
We adapt the ECRB's asymptotic tightness to the case of independent non-identical measurements arising in JCAS, from which the conditional ECRB variant emerges.
Unlike the BCRB used in \cite{xiong2023}, the ECRB enables an exact asymptotic characterization in our setting.
Moreover, we showcase the established fundamental trade-off through two practically-inspired use cases, one involving DoA estimation and another involving OFDM signaling (see Section~\ref{section:examples}).

An interesting feature revealed by our analysis is that all optimal rate-MSE trade-off points are characterized by Gaussian input distributions with different choices of covariance matrices. 
In the terminology of Xiong \emph{et al.} \cite{xiong2023}, the MIMO JCAS setting considered here exhibits only an ST and no DRT, with Gaussian-like communication waveforms also being optimal for sensing.
This is mainly because the sensing channel parameter is fixed.
Arriving at this conclusion, however, is nontrivial, since i.i.d. Gaussian codebook ensembles widely used in standard achievability proofs do not directly apply in our case. Intuitively, i.i.d.-generated codebooks occasionally contain codewords that are unfavorable for sensing.
Alternatively, we rely on \emph{almost-constant covariance codes} \cite{joudeh2023}, which yield uniformly favorable sensing waveforms while asymptotically exhibiting Gaussian-like properties that preserve their communication optimality.
\subsubsection*{Related work} 
A capacity-MSE trade-off, similar to the one we present here, was recently derived by Seo \cite{seo2025a}, albeit in a much more restricted setting.
A major difference is that the approach of Seo is only applicable to discrete memoryless channels (DMCs), where inputs and outputs come from finite sets (despite the parameter being continuous).
In particular, Seo follows the approach developed by Wu and Joudeh in \cite{wu2024,wu2022}, which relies on constant composition codes to prove achievability and type-by-type analysis to prove the converse.
Such an approach is not applicable to Gaussian (or other continuous) channels, and cannot be used to analyze the MIMO JCAS setting at hand.
In contrast, our achievability proof is based on Feinstein's lemma, which is applicable to channels with general alphabets \cite{polyanskiy2025}.
Moreover, our converse proof avoids the type-based argument in \cite{seo2025a} and instead relies on a convexity property of the conditional ECRB, which we also prove in this paper. 
Another smaller difference is that Seo considers a scalar sensing channel parameter only, while we consider a general vector parameter.

The MIMO JCAS channel in the fixed-parameter regime has also been studied in the prior literature, e.g., \cite{fliu2022,hua2024,ren2024,xu2024}. With the exception of \cite{xu2024}, however, these works adopt the classical CRB within a non-Bayesian framework. As noted earlier, they also focus primarily on signal processing and optimization aspects without providing information-theoretic optimality guarantees. 
Moreover, the reliance on single-letter models, prevalent in the wireless communication literature, typically involves an informal notion of empirical input covariance convergence under the generic pretext of ``Gaussian signaling" (see, e.g., \cite[eq.~(3)]{fliu2022} and \cite{hua2024}). We address this issue by adopting a multi-letter perspective and employing almost-constant covariance codes. 

Finally, although Xiong \emph{et al.} \cite{xiong2023} consider the i.i.d. varying-parameter regime, the obtained rate-BCRB trade-off can be applied to the Bayesian fixed-parameter regime we consider in the current paper. As we show, this yields an outer bound that is strictly loose, and hence not achievable, in general. This has also been observed by Seo \cite{seo2025a} in the context of the DMC.  
\subsubsection*{Notation}
We adopt basic notation, which is mostly clear from the context. 
Upper-case bold letters as $\bm{A}$ denote matrices; lower-case bold letters as $\bm{a}$ denote (column) vectors; while $A$ and $a$ are scalars. Calligraphic letters as $\mathcal{A}$ denote sets.
Random variables are denoted as $\mathbf{A}$, $\mathbf{a}$ and $\mathrm{a}$, while $\bm{A}$, $\bm{a}$ and $a$ denote deterministic quantities (e.g., realizations).
\section{Problem Setting}
\label{sec:model}
We consider a vector (i.e., MIMO) JCAS channel comprising an $M$-antenna transmitter, an $L$-antenna receiver (or user), and a $T$-antenna sensor.\footnote{The vector dimension is interpreted as antennas, but can also represent time extension or subcarriers (e.g., OFDM).}
The transmitter sends an encoded signal over the channel with the dual purpose of \emph{communicating} an information message to the user, and enabling \emph{sensing} of a target or the environment by generating back-scattered signals for the sensor.
We assume that the sensor has access to the transmitted encoded signal and can perform coherent sensing. This is the case in monostatic sensing, where the transmitter and sensor are co-located; or in bistatic sensing with a strong link between the transmitter and sensor.
\subsection{Signal model}
We consider a baseband discrete-time model, in which each channel use can be interpreted as one symbol period. 
In the $n$-th use, the transmitter sends a vector channel symbol $\xRvVec_n \in \C^M$ through its $M$ antennas.
The user receives a linearly-transformed noisy vector 
$\yRvVec_n \in \C^L$ given as
\begin{equation}
\label{eq:signal_y}
    \yRvVec_n = \HMat \xRvVec_n + \bm{\omega}_{n}
\end{equation}
where $\HMat \in \C^{L \times M}$ is the channel matrix between the transmitter and the user and $\bm{\omega}_{n} \sim \CN(\mathbf{0}, \sigma^2_{\bm{\omega}} \bm{I})$ is a zero-mean circularly symmetric complex Gaussian noise vector, independent over channel uses $n$.
We refer to $\HMat$ as the \emph{user channel matrix}. This remains fixed over channel uses in the setting we consider in this paper, and it is assumed to be known to both the transmitter and the user.
An underlying assumption here is that initial access, including channel training, between the transmitter and the user has already been performed. 

\begin{figure}[t]
    \centering
    \includegraphics[width=0.7\linewidth]{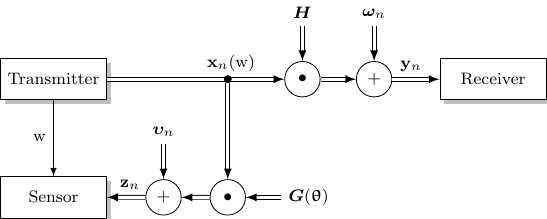}
    \caption{The considered MIMO JCAS system model.}
    \label{fig:model}
\end{figure}

The sensor receives a noisy echo signal ${\zRvVec_n \in \C^T}$ through a back-scatter channel modeled by
\begin{equation}
\label{eq:signal_z}
    \zRvVec_n = \GMat(\thetaRvVec) \, \xRvVec_n + \bm{\upsilon}_{n}
\end{equation}
where $\GMat(\thetaRvVec) \in \C^{T \times M}$ is the channel response matrix from the transmitter to the target (or environment) and back to the sensor, and $\bm{\upsilon}_n \sim \CN(\mathbf{0}, \sigma^2_{\bm{\upsilon}} \bm{I})$ is the corresponding additive noise vector.
We refer to $\GMat(\thetaRvVec)$ as the \emph{sensor channel matrix}. It is a function of a real
\emph{target parameter vector} $\thetaRvVec \in \mathbb{R}^K$, drawn from a continuous distribution with probability density function $f_{\thetaRvVec}(\bm{\theta})$ on $\mathbb{R}^K$.
Here, $K$ denotes the number of parameters to be estimated.
Furthermore, $\bm{\uptheta}$ can represent a single point-target parameter vector or multiple scattering parameters, possibly across multiple targets. 
The sensor is interested in estimating the target parameter vector, as it contains relevant information about the target or environment. 
We assume that the mapping $\GMat: \mathbb{R}^K \to \C^{T \times M}$ from the target parameter vector to the sensor channel matrix is differentiable and invertible (almost surely on the support of $\thetaRvVec$).
As we shall see in Section~\ref{section:examples}, where we discuss particularized examples of this model, these assumptions are justifiable in many practical settings. 
An illustration of our model is shown in Fig.~\ref{fig:model}.

Note that while $\thetaRvVec$ (and therefore $\GMat(\thetaRvVec)$) is drawn at random, it remains fixed throughout the transmission block in the setting we consider. 
This represents scenarios where the target parameter changes at a much slower time scale compared to channel uses within a single transmission block, which is the case in many practical scenarios, as explained earlier.

Transmission occurs over a block of $N$ channel uses, where a signal (i.e., a sequence of vector channel symbols) given by $\xVec^N \triangleq \xVec_1, \xVec_2, \dots, \xVec_N$ is sent. As we specify shortly, the transmitted signal $\xVec^N$ is a codeword selected from a codebook, and encodes a message intended for the user. It also serves the additional purpose of probing to enable parameter estimation at the sensor. 
The transmitted signal is subject to an average (over symbols) power constraint expressed by
\begin{equation}
    \label{eq:power_const}
    \frac{1}{N} \sum_{n=1}^N \lVert \xVec_n \rVert^2 \leq P
\end{equation}
where $P$ is the average transmit power budget.
\subsection{Encoding, decoding, and estimation}
We now formally describe a general JCAS scheme for the above setting, which involves encoding at the transmitter, decoding at the user, and estimation at the sensor.
\subsubsection{Encoding}
In a given transmission block of length $N$, the transmitter wishes to communicate a message $\mathrm{w}$ to the user, where $\mathrm{w}$ is drawn uniformly at random from a message set of $M_N$ indices given by $\mathcal{M}_N \triangleq \{1,2,\dots,M_N\}$.
To this end, an encoded signal (or codeword) $\xVec^N(w) = \xVec_1(w),\ldots,\xVec_N(w)$ is sent. The set of all codewords, i.e., codebook, is given by
\begin{equation*}
    \bm{\mathcal{C}}_N \triangleq \left\{ \xVec^N(1), \xVec^N(2), \dots, \xVec^N(M_N) \right\}.
\end{equation*}
Each codeword in the codebook must satisfy the power constraint in (\ref{eq:power_const}).
Moreover, since $\mathrm{w}$ is uniform, then the random transmitted signal $\xRvVec^N \triangleq \xVec^N(\mathrm{w})$ is uniform on $\bm{\mathcal{C}}_N$.
Note that the signals received by the user and the sensor depend on $\mathrm{w}$ through $\xVec^N(\mathrm{w})$; hence, conditioning on $\mathrm{w} = w$ in what follows is often replaced with conditioning on $\xRvVec^N = \xVec^N(w)$.
\subsubsection{Decoding}
The user receives a random signal given by an $N$-sequence of vectors $\yRvVec^N \triangleq \yRvVec_1, \yRvVec_2, \dots, \yRvVec_N$ from which it produces a decoded message $\hat{w}(\yRvVec^N) \in \mathcal{M}_N$. 
Recall the assumption that the user has access to its channel matrix $\HMat$, on which the decoder may depend.
Given that $\mathrm{w} = w$ is sent, and hence $\xRvVec^N  = \xVec^N(w)$ is transmitted, the probability of decoding error is 
\begin{equation}
    \varepsilon_{N}(w) \triangleq \Prb \left[ \hat{w}(\yRvVec^N) \neq \mathrm{w} \mid \xRvVec^N  = \xVec^N(w) \right].
\end{equation}
The \emph{communication reliability} is measured by the maximal probability of decoding error
\begin{equation}
    \label{eq:pe}
    \varepsilon_{N} \triangleq \max_{w \in \mathcal{M}_N}  \varepsilon_{N}(w).
\end{equation}
We shall refer to this as the \emph{error probability} for short.
The maximum over all messages in \eqref{eq:pe} reflects an underlying assumption that all messages are equally important, i.e., there is no reason to distinguish between different message reliabilities.  It is sometimes technically more convenient to work with the average (over messages) decoding error probability $\varepsilon_{N}^{\mathrm{av}} = \frac{1}{M_N}\sum_{w \in \mathcal{M}_N} \varepsilon_{N}(w)$.
However, as is often the case in asymptotic analysis of point-to-point communication channels, capacity results remain unchanged regardless of whether we choose $\varepsilon_{N}$ or $\varepsilon_{N}^{\mathrm{av}}$ as the measure of decoding error probability. This is also the case for the results we present in this paper.
\subsubsection{Estimation}
\label{subsubsec:estimation}
The sensor receives the random signal given by $\zRvVec^N \triangleq \zRvVec_1, \zRvVec_2, \dots, \zRvVec_N$, and with knowledge of the codeword $\xVec^N(w)$, it produces an estimate $\hat{\thetaVec}_N(\zRvVec^N,\xVec^N(w))$ of the random target parameter vector $\thetaRvVec$. Note that the estimation function  $\hat{\thetaVec}_N: \mathbb{C}^{T \times N} \times \bm{\mathcal{C}}_N \to \mathbb{R}^K$ itself is a deterministic mapping, while the estimate $\hat{\thetaVec}_N(\zRvVec^N,\xVec^N(w)) $ is a random variable due to the randomness of the observation $\zRvVec^N$.
For brevity, we sometimes write $\hat{\thetaVec}_N(\zRvVec^N) $ as $\xVec^N(w)$ is given.

A quadratic loss criterion is used to measure the estimation error, leading to an MSE-based estimation risk described next.
To formulate the estimation risk, we shall first define the MSE matrix.
Given  an estimator $\hat{\thetaVec}_N$ and that $\xVec^N(w)$ is sent, the MSE matrix is defined as
\begin{equation}
\label{eq:MSE_matrix}
\mathbf{\Sigma} \left(\xVec^N(w) , \hat{\thetaVec}_N \right) \triangleq
    \E \left[ \left[ \thetaRvVec -\hat{\thetaVec}_N(\zRvVec^N) \right] \left[ \thetaRvVec -\hat{\thetaVec}_N(\zRvVec^N) \right]^\trans  \: \Big| \: \xRvVec^N = \xVec^N(w)\right]
\end{equation} 
where the expectation is with respect to $\thetaRvVec $ and $\zRvVec^N$, given $ \xRvVec^N = \xVec^N(w)$.\footnote{Our choice of MSE is sometimes referred to as the Bayesian MSE \cite{kay1993}, since it involves an expectation over the prior of $\thetaRvVec$.} 
The diagonal elements of $\mathbf{\Sigma}(\xVec^N(w) , \hat{\thetaVec}_N)$ represent the MSEs of individual target parameters in $\thetaRvVec$. 
The error in estimating the target parameter vector, given codeword $\xVec^N(w)$, is measured in terms of the sum-MSE risk (or \emph{MSE  risk} for short), defined as 
\begin{equation}
\label{eq:MSE_w}
    \xi_N(w) \triangleq  \trace{ \mathbf{\Sigma} \left(\xVec^N(w) , \hat{\thetaVec}_N \right) }.
\end{equation}
The \emph{sensing risk} is chosen to be the \emph{maximal  MSE  risk} (maximal over messages) defined as 
\begin{equation}
\label{eq:max_MSE}
    \xi_N \triangleq \max_{w \in \mathcal{M}_N} \xi_N(w).
\end{equation}
The maximum over messages in \eqref{eq:max_MSE} ensures a uniform sensing risk across different messages, or codewords in the codebook. This reflects a natural operational requirement that the sensing performance should be independent of which message is being communicated, or equivalently, which codeword is being transmitted.
Such a formulation gives rise to the interesting design problem of finding \emph{good} channel codebooks with the important property that all codewords also constitute \emph{equally good} probing signals for the sensing task.

A less stringent measure of sensing risk is the average (over messages) MSE risk, defined as $ \xi_N^{\mathrm{av}} \triangleq \frac{1}{M_N}\sum_{w \in \mathcal{M}_N} \xi_N(w)$. This, however, does not provide the same sensing performance guarantees. Interestingly, as we shall see further on, regardless of whether we choose $\xi_N$ or $\xi_N^{\mathrm{av}}$, the fundamental asymptotic trade-off we establish in the next section remains unchanged.
\begin{remark}
\label{remark:WMSE}
In many practical settings, individual target parameters may carry different importance. For such scenarios, we may introduce the weights $a_1,a_2,\ldots,a_K$, where $a_i \geq 0$ signifies the ``importance'' of the $i$-th parameter in $\thetaRvVec$.
Weights are assumed to be fixed and known beforehand. 
In this case, we modify the MSE in \eqref{eq:MSE_w} into a weighted MSE (WMSE)
\begin{align}
\label{eq:WMSE_w}
    \xi_N(w) &\triangleq \sum_{i = 1}^K a_i \left[ \mathbf{\Sigma} \left(\xVec^N(w) , \hat{\thetaVec}_N \right) \right]_{ii} \\
    &= \trace{ \bm{A} \mathbf{\Sigma} \left(\xVec^N(w) , \hat{\thetaVec}_N \right) }
\end{align}
where  $\bm{A} \triangleq \diag(a_1,a_2,\ldots,a_K)$ is a diagonal matrix induced by the weights.
The WMSE sensing risk is then obtained from \eqref{eq:WMSE_w} by taking the maximum or average over messages. 
For ease of exposition, we will present our results in terms of the unweighted MSE risk. Nevertheless, all results and proofs can be easily extended to account for the weighted case.
\end{remark}
\subsection{Asymptotic rate-MSE trade-off}
A JCAS scheme, as defined above, is referred to as an $(N, M_N, \varepsilon_N, \xi_N)$-scheme if it has a codebook $\bm{\mathcal{C}}_N$ of $M_N$ length-$N$ codewords, a decoder with maximal error probability $\varepsilon_N$, and an estimator with maximal MSE risk $\xi_N$.
We are interested in characterizing the fundamental performance limits of such schemes in the asymptotic regime of $N \to \infty$.

As the block-length $N$ grows, it is known from the channel coding theorem that the size of the message set $M_N$ can be made to grow exponentially in $N$ while guaranteeing reliable decoding, i.e., $\varepsilon_N \to 0$.
Therefore, communication efficiency is often measured by the \emph{rate} $\frac{1}{N} \log M_N$, which is understood as the number of communicated bits per channel use.

On the other hand, we expect the sensing risk in \eqref{eq:max_MSE} to decay to zero as $N$ grows large at a rate of $O(1/N)$, since $\thetaRvVec$ is fixed and the measurements are conditionally independent \cite{vanTrees2013}. 
Therefore, an appropriate asymptotic measure of sensing risk is the scaled MSE $N \xi_N$. 
We will refer to the scaled MSE as the \emph{asymptotic MSE}, or \emph{MSE} for short, where it should be clear from the context whether we are referring to $\xi_N$ or the scaled version $N \xi_N$.
We are now in position to introduce the notion of asymptotically achievable rate-MSE pairs.
\begin{defn}
\label{defintion:achievable}
    The rate-MSE tuple $(R,\Delta)$ is said to be achievable if for every $\epsilon > 0$ and $N \geq N_{\epsilon}$ (possibly large and may depend on $\epsilon$), there exists an $(N, M_N, \varepsilon_N, \xi_N)$-scheme with
    \begin{align*}
        \varepsilon_N & \leq \epsilon \\
        \frac{1}{N} \log M_N & \geq R - \epsilon \\
        N \xi_N & \leq \Delta + \epsilon. 
    \end{align*}
The capacity-MSE function $C(\Delta)$ is the maximum rate $R$ for which $(R,\Delta)$ is achievable.
\end{defn}
Stated differently, the above definition says that if a rate-MSE tuple $(R,\Delta)$ is achievable, then there exists an $N$-block scheme for the above setting that can reliably communicate any one of $M_N = 2^{N (R - \epsilon)} $ messages, with error probability $\varepsilon_N = \epsilon$; while simultaneously achieving a sensing risk of $\xi_N = (\Delta + \epsilon) / N$. Note that $\epsilon$ can be made as small as desired by choosing a large-enough block-length $N$.
Our main result is to fully characterize $C(\Delta)$ (see Theorem~\ref{theorem:capacity_mse}), and showcase the trade-off through a couple of canonical examples.
\begin{remark}
\label{remark:asymptotics}
Before we proceed, we justify some of our assumptions in this remark.
First, for the fixed target parameter, this is practically motivated by the fact that many parameters of interest in sensing remain unchanged or change very slightly over a typical transmission block. For
instance, during a transmission period of $0.5$ ms, a fast-traveling car at $100$ km/h moves by just over $1.3$ cm, and hence parameters of interest (e.g., range, velocity, angle) remain almost fixed.

Now moving on to the asymptotic regime ($N \to \infty$), we argue that this does not pose a contradiction when considered together with a fixed parameter. 
For instance, consider a wireless JCAS system operating with $B= 250$ MHz of bandwidth. A transmission block spanning $\tau = 0.5$ ms corresponds to roughly $N=125000$ channel uses (i.e. $N \approx \tau B$ complex baseband samples). Even if we consider smaller bandwidths, shorter intervals, or lower bandwidth utilization due to pulse shaping, we will still obtain a large enough $N$ that justifies asymptotic analysis.
\end{remark}
\section{Main Result}
\label{sec:main}
\subsection{Preliminaries}
To facilitate the presentation of our main result, we start by introducing common information measures, namely the mutual information and the Fisher information.
We also introduce a function of the Fisher information matrix (FIM), related to the ECRB, which will be used to characterize the asymptotic MSE in the main theorem. 
In what follows, we use $f_{\yRvVec|\xRvVec}(\bm{y} | \bm{x})$ to denote the density of $\CN(\bm{H}\bm{x},\sigma^2_{\bm{\omega}}\bm{I})$, i.e., the transition law of the user channel \eqref{eq:signal_y}.
Similarly, $f_{\zRvVec|\xRvVec,\thetaRvVec}(\bm{z} | \bm{x},\bm{\theta})$ denotes the density of $\CN(\bm{G}(\bm{\theta})\bm{x},\sigma^2_{\bm{\upsilon}}\bm{I})$, associated with the sensor channel \eqref{eq:signal_z}.

Now suppose that the random vector $\xRvVec \in \mathbb{C}^M$ is used as input in a single use of the JCAS channel under consideration.
The mutual information between $\xRvVec $ and the user output $\yRvVec$ is
\begin{equation}
\label{eq:MI}
    I(\xRvVec;\yRvVec) \triangleq \E \left[ \log \left( \frac{f_{\yRvVec|\xRvVec}(\yRvVec | \xRvVec)}{f_{\yRvVec}(\yRvVec)} \right) \right]
\end{equation}
where $f_{\yRvVec}(\bm{y})$ is the probability density of the output $\yRvVec$ induced by the input $\xRvVec$. 
Note that the quantity inside the expectation in \eqref{eq:MI} is known as the information density, defined for any given input-output pair of realizations $\xVec$ and $\yVec$ as 
\begin{equation}
\label{eq:ID}
    \imath(\xVec;\yVec) \triangleq \log \left( \frac{f_{\yRvVec|\xRvVec}(\yVec | \xVec)}{f_{\yRvVec}(\yVec)} \right).
\end{equation}
For $\tilde{\xRvVec} \sim \CN(\mathbf{0},\bm{Q})$, the mutual information 
evaluates to 
\begin{equation}
\label{eq:Gaussian_MI}
    I(\tilde{\xRvVec};\yRvVec) = \log \det \left( \bm{I} + \frac{1}{\sigma^2_{\bm{\omega}}}\bm{H}\bm{Q} \bm{H}^{\herm} \right).
\end{equation}
It is well known that \eqref{eq:Gaussian_MI} is non-decreasing, and concave in $\bm{Q} \in \mathbb{S}_{+}^M$, where $\mathbb{S}_{+}^M$ denotes the cone of $M \times M$ complex, symmetric, positive semi-definite matrices.

Now let us condition on  $\thetaRvVec = \bm{\theta}$ and $\xRvVec = \bm{x}$. The FIM associated with estimating $\bm{\theta}$ from the random sensor observation $\zRvVec \sim f_{\zRvVec|\xRvVec,\thetaRvVec}(\bm{z} | \bm{x},\bm{\theta})$ is given by
\begin{equation}
  \label{eq:FIM}
  \FIM(\thetaVec|\bm{x}) \\
  \triangleq \E \left[ \left[ \nabla_{\thetaVec} \ln f_{\zRvVec|\xRvVec,\thetaRvVec}(\zRvVec | \bm{x},\bm{\theta}) \right] \left[ \nabla_{\thetaVec} \ln f_{\zRvVec|\xRvVec,\thetaRvVec}(\zRvVec | \bm{x},\bm{\theta}) \right]^{\trans} \right].
\end{equation}
For our case, the $(i,j)$-th element of $  \FIM(\thetaVec|\bm{x}) $ evaluates to 
\begin{align}
    \left[ \FIM(\thetaVec|\bm{x}) \right]_{ij} 
    &= \frac{2}{\sigma^2_{\bm{\upsilon}}}\Re\left\{ \xVec^\herm\frac{\partial  \GMat(\thetaVec)^\herm}{\partial\theta_i}\frac{\partial  \GMat(\thetaVec)}{\partial\theta_j}\xVec\right\}\\
    \label{eq:FIM_ij}
    &=\frac{2}{\sigma^2_{\bm{\upsilon}}}\Re\left\{ \trace{\frac{\partial  \GMat(\thetaVec)}{\partial\theta_j}\xVec\xVec^\herm\frac{\partial  \GMat(\thetaVec)^\herm}{\partial\theta_i}}\right\}.
\end{align}
A quantity derived from the FIM, which is very relevant to us, is the following:
\begin{equation}
\label{eq:ECRB_matrix}
    \bm{K}(\thetaRvVec; \zRvVec | \xRvVec) \triangleq  \E \left[ \E \left[ \FIM(\thetaRvVec| \xRvVec ) | \thetaRvVec\right]^{-1} \right].
\end{equation}
To parse \eqref{eq:ECRB_matrix}, given $\thetaRvVec = \thetaVec$, we take the expectation $\E \left[ \FIM(\thetaVec|\xRvVec ) \right]$ with respect to a random input $\xRvVec$, which we will refer to as the \emph{conditional FIM}.\footnote{This is similar, in a sense, to the conditional entropy \cite{Cover}, which involves conditioning and then taking an expectation. Note that for fixed $\thetaVec$, the conditional FIM $\E \left[ \FIM(\thetaVec|\xRvVec ) \right]$ appears in the \emph{modified CRB} \cite{dandrea2002}.} We then invert the conditional FIM and take its expectation with respect to the random parameter $\thetaRvVec$.
As shown in Section \ref{sec:proof}, $\bm{K}(\thetaRvVec; \zRvVec | \xRvVec) $ is closely related to the ECRB for estimating $\thetaRvVec$ from the sensor observations with knowledge of the transmitted codeword, and exactly characterizes the asymptotic MSE.
In fact, we can think of \eqref{eq:ECRB_matrix} as a form of \emph{conditional ECRB} matrix. 

Now suppose that we have an input $\tilde{\xRvVec} \sim \CN(\mathbf{0},\bm{Q})$.
We can see from \eqref{eq:FIM_ij} that the conditional FIM $\E \left[ \FIM(\thetaVec| \tilde{\xRvVec} )  \right]$ in this case only depends on the input distribution through $\bm{Q}$. Therefore, we denote $\E \left[ \FIM(\thetaVec| \tilde{\xRvVec} )  \right]$ by $\FIM(\thetaVec|\bm{Q})$, in which the $(i,j)$-th element is given by
\begin{equation}
    \label{eq:EFIM_Q_ij}
    \left[ \FIM(\thetaVec|\bm{Q}) \right]_{ij} =\frac{2}{\sigma^2_{\bm{\upsilon}}}\Re\left\{ \trace{\frac{\partial  \GMat(\thetaVec)}{\partial\theta_j} \bm{Q} \frac{\partial  \GMat(\thetaVec)^\herm}{\partial\theta_i}}\right\}.
\end{equation}
In this case, the conditional ECRB matrix in \eqref{eq:ECRB_matrix} becomes
\begin{equation}
\bm{K}(\thetaRvVec; \zRvVec | \tilde{\xRvVec}) = \E \left[ \FIM(\thetaRvVec|\bm{Q})^{-1} \right].
\end{equation}
As we will see shortly, the asymptotic MSE is characterized in terms of the function 
\begin{equation}
\label{eq:c_function}
    \trace{  \bm{K}(\thetaRvVec; \zRvVec | \tilde{\xRvVec}) } = \trace{ \E \left[ \FIM(\thetaRvVec|\bm{Q})^{-1} \right] }
\end{equation}
which is perhaps not surprising in view of $\bm{K}(\thetaRvVec; \zRvVec | \tilde{\xRvVec})$ being effectively a conditional ECRB matrix. The exact relationship is discussed in more detail in Section \ref{sec:proof}. 
\begin{lemma}
\label{lemma:d_function}
 $\trace{ \E \left[ \FIM(\thetaRvVec|\bm{Q})^{-1} \right] }$ is monotonically non-increasing and convex in $\bm{Q} \in \mathbb{S}_{+}^M$.
\end{lemma}
The above Lemma will be useful for our analysis. Its proof is presented in Appendix \ref{appendix:lemma_proof}.
\subsection{Result and insights}
We are now in position to present the main result.
\begin{theorem}
\label{theorem:capacity_mse}
The capacity-MSE function $C(\Delta)$ is characterized by 
\begin{equation}
\label{eq:C_Delta_function}
\begin{aligned}
    \max_{\bm{Q} \succeq \bm{0} } &  \  \log \det \left( \bm{I} + \frac{1}{\sigma^2_{\bm{\omega}}}\bm{H}\bm{Q} \bm{H}^{\herm} \right) \\ 
    \mathrm{s.t.} & \ \trace{ \E \left[ \FIM(\thetaRvVec|\bm{Q})^{-1} \right]  } \leq \Delta 
    \\ &  \ \trace{ \bm{Q} } \leq P.
\end{aligned}
\end{equation}
\end{theorem}
A detailed proof of Theorem \ref{theorem:capacity_mse} is presented in Section \ref{sec:proof}.
The rest of this section is dedicated to dissecting the above result and deriving some analytical insights.
\subsubsection{Trade-off}
First, we observe that any optimal rate-MSE pair ${(R,\Delta) = (C(\Delta),\Delta)}$, which must necessarily lie on the capacity-MSE curve $C(\Delta)$, is characterized by 
\begin{align}
R = I(\tilde{\xRvVec};\yRvVec) 
\ \ \text{and} \ \ 
\Delta = \trace{  \bm{K}(\thetaRvVec; \zRvVec | \tilde{\xRvVec}) }
\end{align}
for some Gaussian input vector $\tilde{\xRvVec} \sim \CN(\mathbf{0},\bm{Q})$ with an optimized choice of input covariance matrix $\bm{Q} $ that satisfies the power constraint $\trace{ \bm{Q} } \leq P$.
That is, Gaussian input distributions 
are sufficient for characterizing all optimal rate-MSE trade-off points.
In the terminology of \cite{xiong2023}, the optimal characterization in Theorem \ref{theorem:capacity_mse} only exhibits a so-called ST, but no DRT.
This is due to the fact that $\thetaRvVec$, albeit randomly chosen, remains fixed throughout the transmission.
\subsubsection{Almost-constant covariance codes}
Despite the fact that the capacity-MSE function $C(\Delta)$ is characterized using Gaussian input distributions, the common random coding approach based on i.i.d. Gaussian codebook ensembles \cite{Cover, Telatar1999} seems to be insufficient for proving the achievability of $C(\Delta)$. 
This is mainly due to the worst-case nature of the sensing risk in \eqref{eq:max_MSE}.
The intuition here is that an i.i.d. Gaussian process $\CN(\bm{0}, \bm{Q})$, with a covariance matrix that is optimized to achieve a certain rate-MSE performance, will occasionally generate a codeword $\xVec^N$ that is very bad for sensing. 
In such a \emph{non-typical} (yet possible) scenario, we can have $\hat{\bm{Q}}(\bm{x}^N) \notin \bm{\mathcal{B}}_{N,\delta}(\bm{Q})$ for a given $\delta > 0$ such that $\mathrm{tr}(\E [ \bm{J}( \bm{\uptheta}| \hat{\bm{Q}} (\bm{x}^N) )^{-1} ]) \gg \mathrm{tr}( \E [ \bm{J}( \bm{\uptheta} | \bm{Q} )^{-1}  ] )$, which will dominate the maximal MSE risk in \eqref{eq:max_MSE}.
Alternatively, in our achievability proof, we rely on analyzing codebooks in which all codewords have approximately the same sample covariance matrix. In particular, we impose that 
$\frac{1}{N} \sum_{n = 1}^{N} \bm{x}_n(w)\bm{x}_n(w)^{\herm} \approx \bm{Q}$ must hold for all $\bm{x}^N(w) \in \bm{\mathcal{C}}_N$, in a sense that will be made precise in the proof.
This restriction guarantees a uniform MSE risk performance across all codewords (or messages); at the same time, it yields an achievable rate equal to the mutual information $I(\tilde{\xRvVec};\yRvVec)$, which is identical to the one achieved with i.i.d. Gaussian codebook ensembles with covariance $\bm{Q}$.

Note that in previous works, it is commonly assumed that
$\frac{1}{N} \sum_{n = 1}^{N} \bm{x}_n(w)\bm{x}_n(w)^{\herm} \approx \bm{Q}$ holds under ``Gaussian signaling" as $N$ grows large \cite{fliu2022,hua2024}. The almost-constant covariance codes framework adopted here can be viewed as a rigorous formalization of this assumption.
\subsubsection{Analytical properties}
\eqref{eq:C_Delta_function} constitutes a convex optimization problem, since the objective to be maximized is concave in $\bm{Q}$, while the constraints are convex and linear.
The convex constraint function $\trace{ \E \left[ \FIM(\thetaRvVec|\bm{Q})^{-1} \right]  } $, however, involves an expectation with respect to the prior of $\thetaRvVec$.
Hence, solving \eqref{eq:C_Delta_function} may require approximating this expectation by its sample average. 
In the special case where $\bm{G}(\thetaRvVec)$ is a linear map, the dependency of $\FIM(\thetaRvVec|\bm{Q})$ on the prior washes out, and the problem becomes simpler.
This is seen in Section \ref{section:examples}, where we investigate case studies and present numerical examples. 

It is also useful to note that the function $C(\Delta)$, as described in \eqref{eq:C_Delta_function}, is continuous, non-decreasing, and concave in $\Delta$ (see Appendix \ref{appendix:subsec_properties_C}).
Consequently, the rate increase is expected to slow down as the MSE grows, and exhibit a ``diminishing returns'' behavior due to concavity. This trend is indeed confirmed by the examples in Section \ref{section:examples}.
\subsubsection{BCRB-based bound}
\label{sec:bcrb_comparison}
When proving a converse for $C(\Delta)$, instead of the ECRB-based approach we follow in Section \ref{subsec:converse}, the BCRB (i.e., van Trees inequality) \cite{vanTrees2013} can be used to lower bound the MSE. 
This leads to an upper bound on the capacity-MSE function given by
\begin{equation}
    \label{eq:bcrb_opt}
    \begin{aligned}
        \max_{\bm{Q} \succeq \bm{0} } &  \  \log \det \left( \bm{I} + \frac{1}{\sigma^2_{\bm{\omega}}}\bm{H}\bm{Q} \bm{H}^{\herm} \right) \\ 
        \mathrm{s.t.} & \ \trace{ \E \left[ \FIM(\thetaRvVec|\bm{Q}) \right]^{-1}  } \leq \Delta
        \\ &  \ \trace{ \bm{Q} } \leq P
    \end{aligned}
\end{equation}
which we denote by $C^{\mathrm{ub}}(\Delta)$ (see Appendix \ref{app:C} for details).
Note that the prior information term in the BCRB vanishes since we scale by $N$ and take $N \to \infty$.
It is readily seen that 
\begin{equation}
    \label{eq:bound_comparison}
    \trace{\E \left[ \FIM(\thetaRvVec | \bm{Q}) \right]^{-1}} \leq \trace{ \E \left[ \FIM(\thetaRvVec | \bm{Q})^{-1} \right]}
\end{equation}
which holds due to Jensen's inequality and convexity of the matrix inverse. In fact, due to strict convexity, equality in \eqref{eq:bound_comparison} holds only if $\FIM(\thetaVec|\bm{Q})$ is invariant under $\thetaVec$.
From \eqref{eq:bound_comparison}, it is clear that \eqref{eq:bcrb_opt} gives an upper bound on $C(\Delta)$, which is not achievable in general.
To illustrate this relation, we include the BCRB-based upper bound in our case studies in Section~\ref{section:examples}. 
\begin{remark}
In \cite{xiong2023}, Xiong \emph{et al.} used the Miller-Chang type BCRB \cite{Miller1978} as a proxy for the MSE, where the random input signal $\xRvVec^N$ is treated as a so-called ``nuisance parameter". Adapting this approach to our setting leads to the same upper bound in \eqref{eq:bcrb_opt}, see Appendix \ref{app:C}.
\end{remark}
\section{Proof of Theorem \ref{theorem:capacity_mse}}
\label{sec:proof}
This section is dedicated to proving Theorem \ref{theorem:capacity_mse}.
We start by presenting essential preliminaries on Bayesian parameter estimation, which will be used further on in our proof.
\subsection{Bayesian MMSE estimation and ECRB}
Suppose that a codeword $\xVec^N \in \bm{\mathcal{C}}_N$ is sent (we suppress the dependency on $w$ in this part) and $\zVec^N$ is observed by the sensor.
The optimal estimator that minimizes the MSE risk in \eqref{eq:MSE_w} is the minimum MSE (MMSE) estimator, which is given by the conditional mean
\begin{equation}
    \hat{\thetaVec}_N^{\star}(\zVec^N, \xVec^N) = \E \left[ \thetaRvVec | \zRvVec^N = \zVec^N, \xRvVec^N = \xVec^N\right].
\end{equation}
The corresponding MSE matrix satisfies 
\begin{equation}
\label{eq:MMSE_matrix}
 \mathbf{\Sigma} \left(\xVec^N , \hat{\thetaVec}_N^{\star} \right)  \preceq  \mathbf{\Sigma} \left(\xVec^N , \hat{\thetaVec}_N \right) 
\end{equation}
compared to any estimator $\hat{\thetaVec}_N$, where $\mathbf{\Sigma} \left(\xVec^N , \hat{\thetaVec}_N^{\star} \right)$ is known to be equal to the conditional covariance
matrix of $\thetaRvVec$ given $\zRvVec^N$ and $\xRvVec^N = \xVec^N$ \cite{kay1993,vanTrees2013}.
Note that \eqref{eq:MMSE_matrix} directly implies
\begin{equation}
\trace{ \bm{A} \mathbf{\Sigma} \left(\xVec^N , \hat{\thetaVec}_N^{\star} \right) } \leq \trace{ \bm{A} \mathbf{\Sigma} \left(\xVec^N , \hat{\thetaVec}_N \right) }
\end{equation} 
for any non-negative diagonal weighting matrix $\bm{A}$, and hence optimality is maintained for the WMSE risk in Remark \ref{remark:WMSE}.
We are interested in the asymptotic MSE performance.

By extending \eqref{eq:FIM}, let $\bm{J}(\thetaVec | \xVec^N) $ be the FIM associated with estimating $\thetaRvVec = \bm{\theta}$ from the random sequence of sensor observations $\zRvVec^N$ given that 
$\xRvVec^N = \xVec^N$ is transmitted. From the additive property of the FIM under conditionally independent measurements, we have 
\begin{equation}
\label{eq:FIM_N}
    \bm{J}(\thetaVec | \xVec^N) = \sum_{i = 1}^N \bm{J}(\thetaVec | \xVec_i).
\end{equation}
The classical CRB is obtained by inverting the FIM $ \bm{J}(\thetaVec | \xVec^N) $, while the ECRB is obtained by taking the expectation of the CRB with respect to $\thetaRvVec$, and is given by 
\begin{equation}
\bm{K}(\thetaRvVec; \zRvVec^N | \xRvVec^N = \xVec^N) = \E \left[ \FIM(\thetaRvVec|\bm{x}^N)^{-1} \right].
\end{equation}

Unlike the BCRB \cite{vanTrees2013}, the ECRB does not always yield a lower bound (in the semi-definite sense) on the MSE matrix for every value of $N$.
Nevertheless, in the i.i.d. measurement regime recovered by setting $\xVec_1,\xVec_2,\ldots,\xVec_N = \xVec$ in \eqref{eq:FIM_N}, the ECRB provides an exact characterization of the asymptotic MSE as $N \to \infty$. 
This was stated by Van Trees \emph{et al.} in \cite[Sec.~4.2–4.3]{vanTrees2013} and studied more recently by Aharon and Tabrikian in \cite{aharon2024}. 
In particular, \cite{aharon2024} investigates a weighted variant of the BCRB, showing its validity as a lower bound for all $N$ and its asymptotic tightness, with convergence to the ECRB as $N$ grows large (under regularity conditions).
Next, we adapt this conclusion to the sensing setting at hand, where measurements are independent but non-identical, arising from the fact that the codeword symbols are not identical in general.

To this end, we define the empirical covariance matrix
of an input sequence $\xVec^N$ as 
\begin{equation}
     \hat{\bm{Q}}(\xVec^N) \triangleq \frac{1}{N} \sum_{i =1}^N \xVec_i \xVec_i^{\herm}.
\end{equation}
From \eqref{eq:EFIM_Q_ij}, it is clear that the $N$-letter FIM in \eqref{eq:FIM_N} is equal to a scaled conditional FIM:
\begin{equation}
    \bm{J}(\thetaVec | \xVec^N) =  N \bm{J}(\thetaVec | \hat{\bm{Q}}(\xVec^N)).
\end{equation}
This relation is key to formulating the conditional ECRB tightness result presented next.
\begin{lemma}
\label{eq:lemma_ECRB}
Fix $\epsilon > 0$. There exists a (possibly large) block-length $N_{\epsilon}$ such that 
\begin{equation}
\label{eq:ECRB_Q}
  \trace{\E\left[ \bm{J}(\thetaRvVec | \hat{\bm{Q}}(\xVec^N))^{-1} \right]  }  - \epsilon \leq   N \trace{\mathbf{\Sigma} \left(\xVec^N , \hat{\thetaVec}_N^{\star} \right) } \leq  \trace{ \E\left[ \bm{J}(\thetaRvVec | \hat{\bm{Q}}(\xVec^N))^{-1} \right] }   + \epsilon
\end{equation}
holds for every $N \geq N_{\epsilon}$, provided that some regularity conditions hold.
\end{lemma}
The above lemma follows by slightly adapting existing results for i.i.d. measurements, as shown by Seo \cite{seo2025a} in the scalar parameter case.
In particular, the lower bound in \eqref{eq:ECRB_Q} can be obtained from the weighted BCRB in \cite[Thm.~8]{aharon2024}, which is shown to approach the ECRB as $N$ grows large  (see \cite[Prop.~3]{aharon2024}).
The upper bound in \eqref{eq:ECRB_Q} can be obtained using the sub-optimal maximum likelihood (ML) estimator, whose asymptotic MSE is characterized exactly by the ECRB as discussed in \cite[Sec.~4.2–4.3]{vanTrees2013} (see also \cite[eq.~(12)]{aharon2024}).
The underlying regularity conditions can be found in, e.g., \cite[Thm.~8]{aharon2024}.
While most of these conditions directly hold for Gaussian measurement models with smooth likelihood functions and non-singular FIMs \cite[Thm.~7.3 and App.~7B]{kay1993}, additional ``vanishing'' boundary conditions on the prior distribution $f_{\bm{\uptheta}}(\bm{\theta})$ must be satisfied as well (note that this is also the case for the BCRB \cite{vanTrees2013}).
\subsection{Achievability}
In order to prove achievability, it suffices to show that for an arbitrary fixed covariance matrix $\bm{Q} \succeq 0$ that satisfies $\trace{\bm{Q}} \leq P$, and for every $\epsilon > 0$ and $N$ large enough, there exists an $(N, M_N, \varepsilon_N, \xi_N)$-scheme with $\varepsilon_N \leq \epsilon $ and 
\begin{align}
\label{eq:achievability_rate}
    \frac{1}{N} \log M_N & \geq \log \det \left( \bm{I} + \frac{1}{\sigma^2_{\bm{\omega}}}\bm{H}\bm{Q} \bm{H}^{\herm} \right) - \epsilon \\
\label{eq:achievability_MSE}    
    N \xi_N & \leq  \trace{\E \left[ \FIM(\thetaRvVec|\bm{Q})^{-1} \right]  }  + \epsilon.
\end{align}
By optimizing $\bm{Q}$ and making $\epsilon \to 0$ through $N \to \infty$, we achieve all capacity-MSE trade-off points.
We now break down the achievability proof into three parts as follows.
\subsubsection{Codebook constraint}
For fixed $N$ and codebook $\bm{\mathcal{C}}_{N}$, the sensing risk is dictated by the \emph{worst} codeword, as clearly seen from \eqref{eq:max_MSE}. To guarantee a somewhat uniform MSE across codewords, we impose the constraint that codewords of $\bm{\mathcal{C}}_{N}$ must be drawn from
\begin{equation}
    \label{eq:const_cov}
     \bm{\mathcal{B}}_{N,\delta}(\bm{Q})  \triangleq  \left\{ \xVec^N   \in   \mathbb{C}^{M \times N}  :   \bm{Q}    -   \delta \bm{I}  \preceq \hat{\bm{Q}}(\bm{x}^N)  \preceq \bm{Q}    +   \delta \bm{I}    \right\}
\end{equation}
where $\delta > 0$ is sufficiently small.
All sequences $\bm{x}^N \in \bm{\mathcal{B}}_{N,\delta}(\bm{Q}) $ have an empirical covariance matrix $\hat{\bm{Q}}(\bm{x}^N)$ that is sufficiently close to $\bm{Q}$, in the semi-definite sense. Next, we show that (i) by choosing all codewords from $ \bm{\mathcal{B}}_{N,\delta}(\bm{Q})$, the MSE in \eqref{eq:achievability_MSE} is achieved; and (ii) there exists a codebook with codewords from $  \bm{\mathcal{B}}_{N,\delta}(\bm{Q})$ that achieves the rate in \eqref{eq:achievability_rate}.
\subsubsection{MSE}
Suppose that $\bm{\mathcal{C}}_{N} \subset \bm{\mathcal{B}}_{N,\delta}(\bm{Q})$ and consider the codeword $\xVec^N(w) \in \bm{\mathcal{C}}_{N}$.
Assuming MMSE estimation, and starting from the upper bound in \eqref{eq:ECRB_Q}, we obtain
\begin{align}
N \xi_N(w) & \leq \trace{
  \E\left[ \bm{J}(\thetaRvVec | \hat{\bm{Q}}(\xVec^N(w)))^{-1} \right] }   + \epsilon' \\
  \label{eq:MSE_UB_1}
  & \leq \trace{
  \E\left[ \bm{J}(\thetaRvVec | \bm{Q} - \delta \bm{I} )^{-1} \right] }   + \epsilon' \\
  \label{eq:MSE_UB_2}
  & \leq \trace{ 
  \E\left[ \bm{J}(\thetaRvVec | \bm{Q} )^{-1} \right] } + \delta'  + \epsilon'
\end{align}
where \eqref{eq:MSE_UB_1} is due to monotonicity while \eqref{eq:MSE_UB_2} is due to continuity
(see Lemma \ref{lemma:d_function}).
Note that $\delta'  + \epsilon'$ can be made small by controlling $\delta$ and $N$.
Since \eqref{eq:MSE_UB_2} holds for all $w \in \mathcal{M}_N$, it holds by taking the maximum, from which we obtain \eqref{eq:achievability_MSE}.
\subsubsection{Rate}
We now wish to show that there exists a codebook $\bm{\mathcal{C}}_N \subset \bm{\mathcal{B}}_{N,\delta}(\bm{Q}) $ 
of size $M_N$ that satisfies \eqref{eq:achievability_rate} and has a sufficiently small maximal decoding error probability.

To this end, we invoke a refined version of the channel coding theorem known in the literature as Feinstein's lemma (see \cite{polyanskiy2025} for a comprehensive treatment).
The advantage of using Feinstein's lemma compared to the abundant typicality-based approach \cite{Cover, ElGamal2011} is that it provides guarantees on the maximal (as opposed to average) decoding error probability, and more importantly, it is better suited for handling multi-letter input distributions (non i.i.d. codebooks) and arbitrary codeword constraint sets.
The version we adopt incorporates a constraint set and is referred to in \cite{polyanskiy2025} as the extended Feinstein's lemma.
To this end, recall the information density defined in \eqref{eq:ID}, and let $\imath(\xVec^N;\yVec^N)$ be the corresponding $N$-letter extension.
Moreover, in what follows, let $\tilde{\xRvVec}^N= \tilde{\xRvVec}_{1},\tilde{\xRvVec}_{2},\ldots,\tilde{\xRvVec}_n$ be an i.i.d. sequence of Gaussian vectors, in which every $\tilde{\xRvVec}_n$ is drawn independently from $\CN(\mathbf{0},\bm{Q})$; and $\yRvVec^N$ be a random output sequence induced by the input $\tilde{\xRvVec}^N$.
\begin{lemma}[\textbf{Extended Feinstein's lemma \cite[Thm. 20.7]{polyanskiy2025}}]
\label{lemma:channel_coding_achievability}
Fix $N$ and $\gamma_N> 0$. There exists a codebook $\bm{\mathcal{C}}_N \subset \bm{\mathcal{B}}_{N,\delta}(\bm{Q}) $ of size $M_N$ and maximal decoding error probability $\varepsilon_N$ that satisfies
\begin{equation*}
    \varepsilon_N \times \Prb[\tilde{\xRvVec}^N \in \bm{\mathcal{B}}_{N,\delta}(\bm{Q})] \leq \Prb \left[ \imath(\tilde{\xRvVec}^N; \yRvVec^N) \leq \log \gamma_N \right] + \frac{M_N}{\gamma_N}.
\end{equation*}
\end{lemma}
Note that since $\tilde{\xRvVec}^N$ is i.i.d. and the channel is memoryless, the information density in our case is additive, and for every realization $(\tilde{\xVec}^N,\yVec^N)$ of $(\tilde{\xRvVec}^N,\yRvVec^N)$ we have 
\begin{equation}
   \label{eq:ID_tensorization}
    \imath(\tilde{\xVec}^N;\yVec^N)  = \sum_{n=1}^N \imath(\tilde{\xVec}_{n};\yVec_n) = \sum_{n=1}^N \log \frac{f_{\yRvVec|\xRvVec}(\yVec_n | \tilde{\xVec}_{n})}{f_{\yRvVec}(\yVec_n )}.
\end{equation}
Moreover, since the mutual information is the expected information density (see \eqref{eq:MI}), then 
\begin{equation}
    \E \left[ \imath(\tilde{\xRvVec}^N; \yRvVec^N) \right] =  I(\tilde{\xRvVec}^N; \yRvVec^N) =  N  I(\tilde{\xRvVec} ; \yRvVec)  
\end{equation}
where ${\tilde{\xRvVec} \sim \CN(\mathbf{0},\bm{Q})} $. 
Going back to Lemma \ref{lemma:channel_coding_achievability}, we set 
\begin{equation*}
   \frac{1}{N} \log \gamma_N = I(\tilde{\xRvVec}; \yRvVec) - \frac{\epsilon}{2} 
   \ \ \text{and} \ \ 
   \frac{1}{N} \log M_N  = I(\tilde{\xRvVec}; \yRvVec) - \epsilon.
\end{equation*}
The error probability bound becomes 
\begin{equation}
\label{eq:error_UB_achievability}
\varepsilon_N \times \Prb[\tilde{\xRvVec}^N \in \bm{\mathcal{B}}_{N,\delta}(\bm{Q})] \leq \Prb \left[ \frac{1}{N} \sum_{n = 1}^N \imath(\tilde{\xRvVec}_n ; \yRvVec_n) \leq I(\tilde{\xRvVec}; \yRvVec) - \frac{\epsilon}{2} \right] + 2^{-N \frac{\epsilon}{2}}.
\end{equation}
Since the sequences involved are i.i.d., by the weak law of large numbers (WLLN), we get
\begin{equation}
     \Prb  \left[ \frac{1}{N} \sum_{n = 1}^N \imath(\tilde{\xRvVec}_n ; \yRvVec_n) \leq I(\tilde{\xRvVec}; \yRvVec) - \frac{\epsilon}{2} \right]  \to 0  \ \ \text{as} \ \ N \to \infty.
\end{equation} 
Moreover, we have 
$2^{-N \frac{\epsilon}{2}} \to 0$ as $N \to \infty$. Therefore, the right-hand side of \eqref{eq:error_UB_achievability} approaches zero as $N$ grows large.
The following lemma is key to completing the proof. 
\begin{lemma}
\label{lemma:covariance_concentration}
Let $\tilde{\xRvVec}^N $ be an i.i.d. sequence of Gaussian vectors with distribution $\CN({\bm{0}, \bm{Q}})$. Then
\begin{equation}
    \label{eq:covariance_concentration}
    \Prb[\tilde{\xRvVec}^N \in \bm{\mathcal{B}}_{N,\delta}(\bm{Q})] \to 1 \ \ \text{as} \ \ N \to \infty.
\end{equation}
\end{lemma}
In words, the above lemma says that an i.i.d.  $\CN(\mathbf{0},\bm{Q})$ sequence will, with high probability, have an empirical covariance matrix that is close to $\bm{Q}$ in the positive semi-definite sense. A proof is presented in Appendix~\ref{app:B}.
Putting everything together, we conclude that $\varepsilon_N \leq \epsilon$ is attainable by making $N \to \infty$ large enough, which concludes the proof of achievability.
\subsection{Converse}
\label{subsec:converse}
Here we show that any achievable pair $(R,\Delta)$ must satisfy $R \leq C(\Delta)$, where $C(\Delta)$ is as characterized in Theorem~\ref{theorem:capacity_mse}. 
For this purpose, recall from Definition~\ref{defintion:achievable} that if $(R,\Delta)$ is achievable, then for large enough $N$, there exists an $(N, M_N, \varepsilon_N, \xi_N)$-scheme with $\varepsilon_N \leq \epsilon $, $\frac{1}{N} \log M_N \geq R - \epsilon$ and $N \xi_N \leq \Delta + \epsilon$. We now 
bound the sensing risk as follows
\begin{align}
 \Delta  + \epsilon
&\geq N \max_{w \in \mathcal{M}_N}\xi_N(w)   \\
& \geq  
\label{eq:MSE_LB_1}
\frac{N}{M_N} \sum_{w \in \mathcal{M}_N}\xi_N(w) \\
\label{eq:MSE_LB_2}
& \geq \frac{1}{M_N} \sum_{w \in \mathcal{M}_N} \trace{ \E\left[ \bm{J}(\thetaRvVec | \hat{\bm{Q}}(\xVec^N(w)))^{-1} \right] }   - \epsilon \\
\label{eq:MSE_LB_3}
& \geq \trace{  \E  \left[ \bm{J}  \left(  \thetaRvVec \, \Bigg| \,  \frac{1}{M_N}  \sum_{w \in \mathcal{M}_N}  \hat{\bm{Q}}(\xVec^N(w))   \right)^{-1} \right] }    -  \epsilon
\end{align}
where \eqref{eq:MSE_LB_1} is since the maximum (over messages) is lower bounded by the average, \eqref{eq:MSE_LB_2} is from the lower bound in \eqref{eq:ECRB_Q}, and \eqref{eq:MSE_LB_3} is due to the convexity of \eqref{eq:c_function} and Jensen's inequality (see Lemma \ref{lemma:d_function}).
Next, we denote the empirical covariance matrix average over messages by 
\begin{equation}
   \bar{\bm{Q}} \triangleq  \frac{1}{M_N} \sum_{w \in \mathcal{M}_N} \hat{\bm{Q}}(\xVec^N(w))
\end{equation}
which clearly must also satisfy $  \bar{\bm{Q}}  \succeq \mathbf{0}$ and $\trace{ \bar{\bm{Q}}} \leq P$.
The bound in \eqref{eq:MSE_LB_3} is rewritten as 
\begin{equation}
\label{eq:delta_converse}
     \Delta \geq \trace{ \E\left[ \bm{J}(\thetaRvVec | \bar{\bm{Q}})^{-1} \right] }   - 2 \epsilon.
\end{equation}
We proceed to bound the rate. To this end, we invoke Fano's inequality \cite{Cover, ElGamal2011} and obtain
\begin{align}
R - \epsilon& \leq \frac{1}{N} \log M_N \\
\label{eq:rate_UB_converse}
    &  \leq \frac{1}{N} I(\xRvVec^N ; \yRvVec^N) + \frac{1}{N} +   \frac{\varepsilon_N^{\mathrm{av}}}{N}\log M_N
\end{align}
where the term $ \frac{1}{N} +   \frac{\varepsilon_N^{\mathrm{av}}}{N}\log M_N $ vanishes as $N$ grows large, and hence we bound it by $\epsilon$ from now on.
Moreover, it is worthwhile noting that Fano's inequality yields a bound in terms of $\varepsilon_N^{\mathrm{av}}$. This is, however, valid for our purpose since $\varepsilon_N^{\mathrm{av}} \leq \varepsilon_N$.

The next step is to bound the normalized multi-letter mutual information term $\frac{1}{N} I(\xRvVec^N ; \yRvVec^N) $. To this end, it helps to remember that the random input sequence $\xRvVec^N = \bm{x}^N(\mathrm{w})$ is uniformly distributed on the codebook $\bm{\mathcal{C}}_N$.
Using the fact that the channel is memoryless, we proceed as 
\begin{align}
 \frac{1}{N} I(\xRvVec^N ; \yRvVec^N) 
& \leq \frac{1}{N} \sum_{n = 1}^N I(\xRvVec_n ; \yRvVec_n) \\
\label{eq:rate_UB_1}
    & \leq \frac{1}{N}  \sum_{n = 1}^N \log \det  \left(  \bm{I}  +  \frac{1}{\sigma^2_{\bm{\omega}}}\bm{H} \E  \left[ \xRvVec_n  \xRvVec_n ^{\herm} \right] \bm{H}^{\herm}  \right)  \\
\label{eq:rate_UB_2}    
    & \leq \log \det  \left(  \bm{I}  +  \frac{1}{\sigma^2_{\bm{\omega}}}\bm{H} \frac{1}{N}  \sum_{n = 1}^N  \E  \left[ \xRvVec_n  \xRvVec_n ^{\herm} \right]  \bm{H}^{\herm}  \right) .
\end{align}
The inequality in \eqref{eq:rate_UB_1} follows from the extremal property of Gaussian inputs under a covariance (or correlation) matrix constraint \cite[Sec.~2.2]{ElGamal2011}, which leads to $I(\xRvVec_n ; \yRvVec_n)$ being maximized by an input distribution $\CN(\mathbf{0},\E \left[ \xRvVec_n  \xRvVec_n ^{\herm} \right])$.
\eqref{eq:rate_UB_2} is due to concavity and Jensen's inequality.

Using the fact that $ \xRvVec^N = \bm{x}^N(\mathrm{w})$ is uniformly distributed on the codebook, we observe that
\begin{align*}
    \frac{1}{N} \sum_{n = 1}^N  \E \left[ \xRvVec_n  \xRvVec_n^{\herm} \right]  = \frac{1}{M_N} \sum_{w\in \mathcal{M}_N} \frac{1}{N} \sum_{n = 1}^N  \xVec_n(w)  \xVec_n(w)^{\herm} = \bar{\bm{Q}}.
\end{align*}
Plugging everything back into \eqref{eq:rate_UB_converse}, we obtain
\begin{align}
    R & \leq \log \det \left( \bm{I} + \frac{1}{\sigma^2_{\bm{\omega}}}\bm{H}  \bar{\bm{Q}} \bm{H}^{\herm} \right)  + 2 \epsilon \\
    & \leq C(\Delta + 2 \epsilon)  + 2\epsilon
\end{align}
where the last inequality is obtained by maximizing over $\bar{\bm{Q}} $ subject to the MSE constraint in \eqref{eq:delta_converse}, and the power constraint. 
Taking $\epsilon \to 0$ as $N \to \infty$, we obtain the desired result.
\begin{remark}
    It can be seen from our converse proof that we bounded the maximal (over messages) MSE risk $\xi_N$ in terms of its average counterpart $\xi_N^{\mathrm{av}}$ in \eqref{eq:MSE_LB_1}; and the maximal decoding error probability  $\varepsilon_N$ in terms of its average counterpart $\varepsilon_N^{\mathrm{av}}$ in \eqref{eq:rate_UB_converse}. These are, of course, valid steps as they are applied in the converse part only. More importantly, they imply that the characterization obtained in Theorem \ref{theorem:capacity_mse} remains valid if we adopt less stringent average risk and error measures.
\end{remark}
\section{Case Studies}
\label{section:examples}
\subsection{JCAS with DoA Estimation}
In our first case study, we consider scenarios where the transmitter is communicating with a single antenna user, i.e., $L=1$; while estimating the DoA of a passive point target in its range.
The transmitter and the sensor are equipped with uniform linear arrays (ULAs) of $M$ and $T$ antennas, respectively. Assuming half-wavelength element spacing for both arrays, the steering vectors at the transmitter and at the sensor are given by
\begin{align}
\label{eq:steering_vec_tx}
    \bm{v}_M(\theta)
    &\triangleq \begin{bmatrix}
        1 \; e^{\imagUnit \pi \sin(\theta)} \; e^{\imagUnit 2 \pi \sin(\theta)} \; \dots \; e^{\imagUnit (M-1) \pi \sin(\theta)}
    \end{bmatrix}^{\trans} \\
    \bm{v}_T(\theta)
    &\triangleq \begin{bmatrix}
        1 \; e^{\imagUnit \pi \sin(\theta)} \; e^{\imagUnit 2 \pi \sin(\theta)} \; \dots \; e^{\imagUnit (T-1) \pi \sin(\theta)}
    \end{bmatrix}^{\trans} 
\end{align}
respectively.
We adopt a line-of-sight, single-ray model for the communication channel as
\begin{equation}
    \bm{H} = \beta \, \bm{v}_M(\phi)^{\herm} \in \mathbb{C}^{1 \times M}
\end{equation}
where $\phi \in (-\frac{\pi}{2}, \frac{\pi}{2})$ denotes the direction of departure (DoD) from the transmitter array towards the user's direction, and $\beta \in \C$ denotes the channel gain. As stated in Section \ref{sec:model}, we assume that the user channel matrix is fixed and known to both the transmitter and the user through, e.g., an initial access phase. For simplicity, we set $\beta=1$. 

As for the sensing channel, this is given as
\begin{equation}
\label{eq:doa_channel}
    \GMat(\uptheta) = \lambda \, \bm{v}_T(\uptheta) \, \bm{v}_M(\uptheta)^{\herm} 
    \in \mathbb{C}^{T \times M}
\end{equation}
where $\uptheta$ is the DoA parameter to be estimated, taking values from $(-\frac{\pi}{2},\frac{\pi}{2})$, and $\lambda \in \C$ is the complex back-scatter channel gain (see, e.g., \cite{fliu2022}, among others). 
For simplicity, we assume that the channel gain is known \textit{a priori} and set $\lambda = 1$.
Furthermore, the DoA and the DoD of the sensing channel coincide in our monostatic setting, as the transmitter and the sensor arrays are co-located and parallel.
$\uptheta$ is drawn at random according to a prior $f_{\uptheta}(\theta)$, yet remains fixed throughout the transmission. 
We examine two different prior distributions specified below.
The procedure and analysis can be extended to the joint estimation of DoA and gain \cite{li2008}. 
\subsubsection*{Capacity-MSE trade-off}
To calculate the conditional FIM in accordance with \eqref{eq:EFIM_Q_ij}, which reduces to a scalar for single-parameter estimation, the first-order derivative of the sensing channel matrix with respect to the target parameter $\theta$ is obtained as 
\begin{equation}
    \frac{\partial \GMat(\theta)}{\partial \theta}
    = \imagUnit \pi \cos(\theta) \left( \bm{D}_T \GMat(\theta) - \GMat(\theta) \bm{D}_M \right)
\end{equation}
where $\bm{D}_T \triangleq \diag(0,\dots,T-1)$ and $\bm{D}_M \triangleq \diag(0,\dots,M-1)$ represent element positions in the sensor and transmitter arrays, respectively.
Defining $\tilde{\bm{G}}(\theta) \triangleq \bm{D}_T \bm{G}(\theta) - \bm{G}(\theta) \bm{D}_M$, we obtain 
\begin{equation}
\label{eq:EFIM_DoA}
    J(\theta|\bm{Q}) = \frac{2 \pi^2}{\sigma^2_{\bm{\upsilon}}} \cos^2(\theta)
    \trace{\tilde{\bm{G}}(\theta) \bm{Q} \tilde{\bm{G}}(\theta)^{\herm} }
\end{equation}
where the operator $\Re\{ \cdot \}$  is dropped since the trace of a Hermitian matrix is real.

From Theorem \ref{theorem:capacity_mse}, the capacity-MSE function $C(\Delta)$ specialized for the present use case is
\begin{equation}
\label{eq:opt_doa}
\begin{aligned}
    \max_{\bm{Q} \succeq \bm{0}} &  \  \log \left( 1 + \frac{1}{\sigma^2_{\bm{\omega}}} \bm{v}_M(\phi)^{\herm} \bm{Q} \bm{v}_M(\phi) \right) \\ 
    \mathrm{s.t.} & \: \ \frac{\sigma^2_{\bm{\upsilon}}}{2 \pi^2} \E \left[ \frac{1}{\cos^{2}(\uptheta) \mathrm{tr}(\tilde{\bm{G}}(\uptheta) \bm{Q} \tilde{\bm{G}}(\uptheta))} \right] \leq \Delta \\ & \: \trace{ \bm{Q} } \leq P.
\end{aligned}
\end{equation}
For $\Delta$ sufficiently large, such that the sensing constraint is inactive, 
the capacity is given by
$\log \left( 1 + {MP}/{\sigma^2_{\bm{\omega}}} \right)$,
where $M$ is the array gain. 
In this case, beamforming along the user's DoD is optimal, i.e. ${\bm{Q} = \frac{P}{M} \bm{v}_M(\phi) \bm{v}_M(\phi)^{\herm}}$.
More generally, by optimizing $\bm{Q}$, which reflects the beamforming characteristics of the transmit array, a clear trade-off between the rate and the MSE is obtained, as demonstrated in our numerical results below.

In addition to $C(\Delta)$, we will also examine the BCRB-based upper bound $C^{\mathrm{ub}}(\Delta)$ given in \eqref{eq:bcrb_opt}. This is obtained by replacing the sensing constraint in \eqref{eq:opt_doa} with
\begin{equation}
\frac{\sigma^2_{\bm{\upsilon}}}{2 \pi^2} \frac{1}{\E [ \cos^{2}(\uptheta) \mathrm{tr}(\tilde{\bm{G}}(\uptheta) \bm{Q} \tilde{\bm{G}}(\uptheta)) ]} \leq \Delta
\end{equation}
which is a looser constraint, as discussed in Section~\ref{sec:bcrb_comparison}.
In the numerical results presented below, we set $T=M=16$ for the number of transmitter and sensor antennas.
The communication SNR is defined as $P/\sigma^2_{\bm{\omega}}$ and set to $15$ dB. Similarly, the sensing SNR is defined as $P/\sigma^2_{\bm{\upsilon}}$, and set to $-25$ dB.
We examine two types of prior distribution for $\uptheta$: uniform and non-uniform.
The uniform case is meant to represent scenarios where no prior information is available; while the non-uniform case represents scenarios where prior information suggests that the target is more likely to be in a certain angular sector, as in, e.g., tracking from a previous estimate. 
\subsubsection*{Tapered uniform prior}
\begin{figure}[t]
\centering
\subfloat[prior distribution
\label{fig:doa_uniform_prior}]{
\hspace{-4mm}
\includegraphics[width=0.90\linewidth]{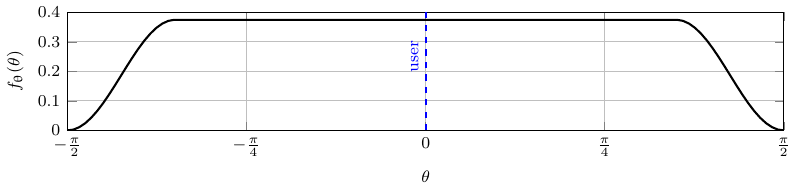}
} \\
\vspace{-4mm}
\subfloat[capacity--MSE trade-off \label{fig:doa_uniform_capacity_mse}]{
\includegraphics[width=0.90\linewidth]{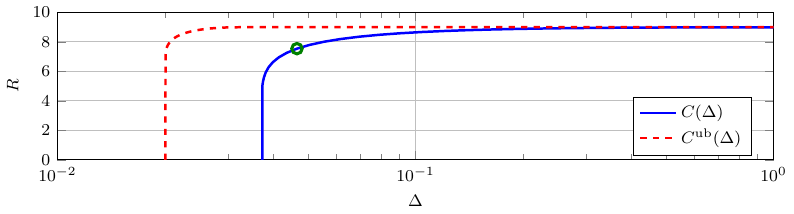}
} \\
\vspace{-4mm}
\subfloat[beamformer evolvement \label{fig:doa_uniform_beamformers}]{
\hspace{-6mm}
\includegraphics[width=0.90\linewidth]{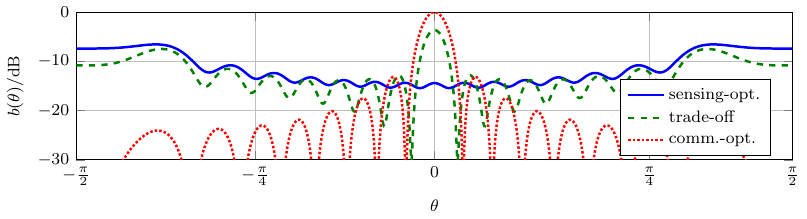}
}
\caption{Capacity–MSE trade-off with a uniform prior in JCAS with DoA estimation.}
\label{fig:DoA_result1}
\end{figure}

Under a uniform prior, one would set $f_{\uptheta}(\theta) = 1/\pi$ for $\theta \in (-\frac{\pi}{2},\frac{\pi}{2})$ and zero otherwise.
However, as highlighted in \cite{vanTrees2002, aharon2024}, Bayesian variants of the CRB (e.g., BCRB and ECRB) are not defined for such a prior, as it violates the regularity conditions.
Therefore, as suggested in \cite{aharon2024}, we adopt a tapered uniform distribution model with a raised-cosine
\begin{equation}
    f_{\uptheta}(\theta) = \frac{1}{s(1+\kappa)}
    \begin{cases}
        1, & |\theta| <  s \kappa \\
        \frac{1}{2} + \frac{1}{2} \cos \left( \frac{ \pi ( |\theta| - s \kappa)}{s(1-\kappa)} \right), & s \kappa \leq |\theta| \leq s \\
        0, & |\theta| > s
    \end{cases}
\end{equation}
where $s \in \R^+$ determines the boundaries of the distribution and $\kappa \in [0,1]$ is the roll-off factor that controls the sharpness at the edges. $\kappa=0$ corresponds to a pure raised-cosine shaped distribution, whereas $\kappa=1$ corresponds to a perfect uniform distribution.
We set $s=\pi/2$ to cover the entire range and $\kappa = 0.7$ to have a tapered uniform distribution. 
Note that since ULAs have lower resolution in the endfire directions, i.e., $\theta \approx \pm \frac{\pi}{2}$, it is reasonable to bias the estimate away from these directions by setting a low prior value.
We set $\phi=0$ for the user channel.
The tapered uniform prior and the user location are shown in Fig.~\ref{fig:doa_uniform_prior}.

Under the above specified assumptions, the capacity-MSE function and its BCRB-based upper bound are shown in Fig.~\ref{fig:doa_uniform_capacity_mse}.
By varying $\Delta$ in \eqref{eq:opt_doa}, and solving the corresponding optimization problem, different rate-MSE trade-off points are obtained. 
Moreover, the looseness of the BCRB-based upper bound compared to the ECRB-based exact characterization is clearly seen.

Next, we also utilize the projection $b(\theta) \triangleq \frac{1}{MP} \bm{v}_M(\theta)^{\herm} \bm{Q} \bm{v}_M(\theta)$ to calculate the normalized beamforming gain in any given angular direction $\theta$, and show the result in Fig.~\ref{fig:doa_uniform_beamformers}.
To illustrate a specific trade-off choice, we select a point on the $C(\Delta)$ curve
indicated with a marker in Fig.~\ref{fig:doa_uniform_capacity_mse}, whose optimal beamforming characteristics are also illustrated in Fig.~\ref{fig:doa_uniform_beamformers}.
While the sensing-optimal beamforming favors a wide scan of the entire range, due to the almost uniform prior; beam steering towards the user's direction gradually turns into the optimal beamformer.
\subsubsection*{Non-uniform prior}
\begin{figure}[t]
    \centering
    \subfloat[prior distribution]{%
        \label{fig:doa_prior_nonuniform}
        \hspace{-5mm}
        \includegraphics[width=0.90\linewidth]{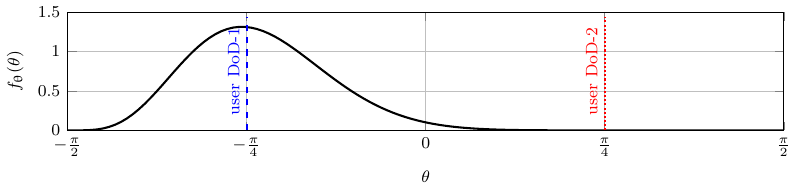}
    }
    \\
    \vspace{-4mm}
    \subfloat[capacity-MSE trade-off]{%
        \label{fig:doa_curve_nonuniform}
        \includegraphics[width=0.90\linewidth]{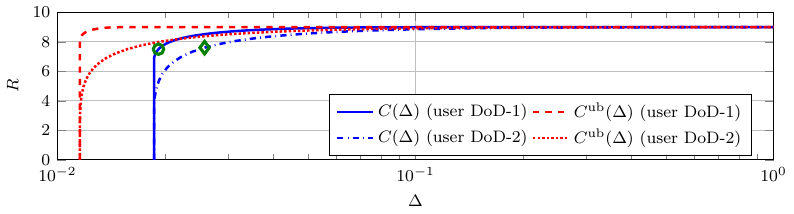}
    }
    \\
    \vspace{-4mm}
    \subfloat[beamformer evolvement]{%
        \label{fig:doa_nonuniform_beamformer}
        \hspace{-5mm}
        \includegraphics[width=0.90\linewidth]{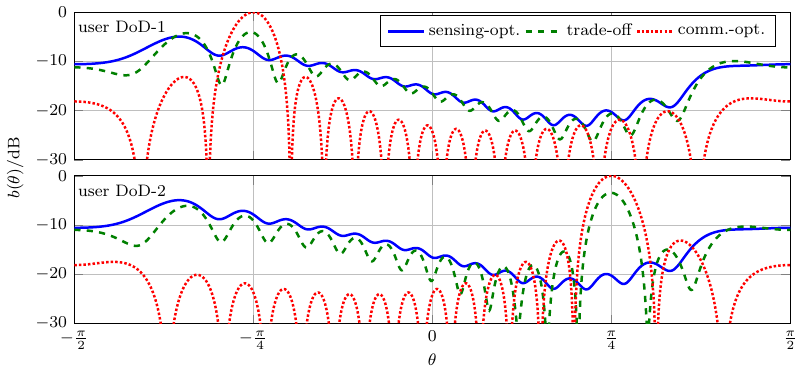}
    }
    \caption{Capacity-MSE trade-off with a non-uniform prior in JCAS with DoA estimation.}
    \label{fig:DoA_result2}
\end{figure}
For this case, we utilize the Beta distribution to model a concentrated prior for which $f_{\uptheta}(\theta) = 0$ when $\theta = \pm \pi/2$ \cite{aharon2024}. In particular, we have
\begin{equation}
f_{\uptheta}(\theta) = \frac{(\theta - \theta_{\min})^{s_1 - 1} (\theta_{\max} - \theta)^{s_2 - 1}}
{(\theta_{\max} - \theta_{\min})^{s_1 + s_2 - 1} B(s_1, s_2)} 
\end{equation}
for $ \theta \in [\theta_{\min}, \theta_{\max}]$ and zero otherwise, where $B(s_1, s_2) \triangleq \int_0^1 t^{s_1 - 1} (1 - t)^{s_2 - 1} \, dt$ is the Beta function, and $s_1$ and $s_2$ are the shape parameters of the distribution.
We set $\theta_{\min} = -\pi/2$ and $\theta_{\max} = \pi/2$ so that the support of the distribution spans the full angular range. The shape parameters are chosen as $s_1 = 5.5$ and $s_2 = 15$, which results in a distribution that is more concentrated around a specific region that peaks around $-\pi/4$.
This selection ensures that the prior still satisfies the regularity conditions while favoring a particular region in the angular domain.
With the non-uniform prior, we consider two scenarios for the communication channel: first, $\phi = -\pi/4$, i.e., the user is located in the region where the target is most likely present, and second, $\phi = \pi/4$, to represent the scenario where the target is well separated from the user in the angular domain.
The prior and the user DoD in the two scenarios are shown in Fig.~\ref{fig:doa_prior_nonuniform}.

The capacity-MSE function and its BCRB-based upper bound are depicted in Fig.~\ref{fig:doa_curve_nonuniform}, for two different user DoD scenarios. 
The optimal beamforming gains for the two scenarios are also shown in Fig.~\ref{fig:doa_nonuniform_beamformer}.
For both user DoD scenarios, a point is selected on the capacity-MSE curve to illustrate the beamforming characteristics of an intermediate trade-off point.

For user DoD-1, since the prior already favors the user direction, the capacity-MSE function exhibits a limited trade-off. 
In this case, the user and the target are highly aligned, and therefore, a $\bm{Q}$ which is good for communication is also good for sensing.
This is in contrast to the user DoD-2 scenario, where the target's prior concentrates further away from the user.
The evolution of beam patterns in Fig.~\ref{fig:doa_nonuniform_beamformer} is also a clear indication of the trade-off in the misaligned case.
\subsection{JCAS with OFDM}
We now consider a cyclic-prefix OFDM scenario, where the vector dimension represents subcarriers instead of antennas.
The system communicates with a user and simultaneously senses the back-scatter channel characterized by the frequency response vector $\thetaRvVec \in \C^K$, where $K$ is the number of subcarriers being utilized.
Assuming that the effect of the cyclic prefix is removed at the receiver, the user channel matrix in this case is given as
\begin{equation}
    \HMat
    =\diag(\FourierMat\bm{h}) \in \mathbb{C}^{K \times K} \label{eq:OFDM_com}
\end{equation}
where $\bm{F} \in \C^{K \times K}$ is the unitary discrete Fourier transform matrix (DFT) and $\bm{h} \in \C^K$ is the impulse response of the communication channel in the time-domain.
We assume that $\bm{H}$ is known to the transmitter and receiver.
Similarly, the sensor channel matrix in \eqref{eq:signal_z} can be written as
\begin{equation}
\GMat(\thetaRvVec) = \diag(\FourierMat \thetaRvVec) \in \mathbb{C}^{K \times K}. \label{eq:OFDM_sens}
\end{equation}
Note that in this use case, we have $M=L=T=K$.
It is also clear that the sensing channel matrix is linear in the target parameter vector $\thetaRvVec$. 
To allow for concise representation, we allow the parameter vector, i.e. $\thetaRvVec$, to be complex-valued. 
The model can be equivalently expressed using a $2K$-dimensional real parameter vector, consistent with \eqref{eq:signal_y} and \eqref{eq:signal_z}. 
For more on the real-valued representation of complex parameters in estimation, see \cite[Ch.~15]{kay1993}.

As we shall see shortly, for this use case, the FIM is invariant under different realizations of the parameter. Hence, the prior density $f_{\thetaRvVec}(\thetaVec)$ does not influence the JCAS trade-off.
\subsubsection*{Capacity-MSE trade-off}
Since the user channel matrix is diagonal, the mutual information in \eqref{eq:Gaussian_MI} is maximized by a diagonal input covariance matrix. Moreover, the conditional FIM and ECRB only depend on the input covariance matrix through its diagonal elements.
To see this, we evaluate the conditional FIM, i.e., \eqref{eq:EFIM_Q_ij}, for this setting. We first calculate the partial derivative of the sensor channel matrix with respect to the $i$-th entry of the parameter as
\begin{equation}
    \frac{\partial \GMat(\theta)}{\partial \theta_i}
    = \diag( \bm{f}_i )
\end{equation}
where $\bm{f}_i$ denotes the $i$-th column of the unitary DFT matrix $\bm{F}$.
From this, we obtain
\begin{align}
    [\FIM(\thetaVec|\bm{Q})]_{ij}
    &=\frac{1}{\sigma^2_{\bm{\upsilon}}} \trace{ \frac{\partial \GMat(\theta)}{\partial \theta_j} \bm{Q} \frac{\partial \GMat(\theta)}{\partial \theta_i} }\\
    &= \frac{1}{\sigma^2_{\bm{\upsilon}}} \trace{ \diag( \bm{f}_j ) \bm{Q} \diag( \bm{f}_i^{\herm} ) }\\
    &= \frac{1}{\sigma^2_{\bm{\upsilon}}} \bm{f}_i^{\herm} \left( \bm{I} \circ \bm{Q} \right) \bm{f}_j
\end{align}
where $\circ$ denotes the Hadamard (element-wise) product between two matrices.
The resulting conditional FIM can be expressed more compactly as
\begin{equation}
    \FIM(\thetaVec|\bm{Q}) = \frac{1}{\sigma^2_{\bm{\upsilon}}} \FourierMat^\herm  (\bm{I}\circ\bm{Q})\FourierMat.
    \label{eq:EFIM_OFDM}
\end{equation}
Note that this also directly follows from the FIM for linear models  \cite[Ex.~15.9]{kay1993}.
From \eqref{eq:EFIM_OFDM}, we see that the conditional FIM does not depend on the off-diagonal terms of $\bm{Q}$. Hence, restriction to diagonal matrices does not influence the sensing performance either.
Therefore, in what follows, we assume without loss of generality that $\bm{Q} = \diag(\bm{\sigma}_{\xRvVec}^2)$, where $\bm{\sigma}_{\xRvVec}^2 \triangleq (\sigma_{\xRvVec,0}^2,\ldots,\sigma_{\xRvVec,K-1}^2)$.

Continuing with a diagonal input covariance,
\eqref{eq:EFIM_OFDM} becomes
\begin{equation}
    \label{eq:OFDM_FIM}
    \FIM(\thetaRvVec|\bm{\sigma}^2_\xRvVec ) = \frac{1}{\sigma^2_{\bm{\upsilon}}} \FourierMat^\herm  \diag(\bm{\sigma}^2_\xRvVec)\FourierMat.
\end{equation}
and the asymptotic MSE is given by
\begin{equation}
  \trace{ \FIM(\thetaRvVec|\bm{\sigma}^2_\xRvVec )^{-1} } 
  =\sigma^2_{\bm{\upsilon}}\sum_{k=0}^{K-1} \frac{1}{\sigma^2_{\xRvVec,k}}. \label{eq:OFDM_MSE}
\end{equation}
On the other hand, the rate expression is easily obtained by specializing \eqref{eq:Gaussian_MI}, leading to the well-known sum-rate expression of parallel Gaussian channels \cite[Sec.~9.4]{Cover}.
Combining this with Theorem \ref{theorem:capacity_mse}, we obtain the 
capacity-MSE trade-off
\begin{equation}
\label{eq:opt_ofdm}
\begin{aligned}
    \max_{\diag(\bm{\sigma}^2_{\xRvVec}) \succeq \bm{0}} &\  \sum_{k=0}^{K-1} \log \left( 1 + \frac{\sigma^2_{\xRvVec,k}}{\sigma^2_{\bm{\omega}}} \vert [\bm{H}]_{kk}\rvert^2 \right) \\ 
    \mathrm{s.t.} & \quad \sigma^2_{\bm{\upsilon}} \sum_{k=0}^{K-1} \frac{1}{\sigma^2_{\xRvVec,k} } \leq \Delta \\
    & \quad \sum_{k=0}^{K-1} \sigma^2_{\xRvVec,k} \leq P
\end{aligned}
\end{equation}
which boils down to power allocation across parallel channels under an ECRB constraint.
\begin{remark}
In this special case, where the sensor observations are a linear function of the parameter vector corrupted by Gaussian noise,
the conditional FIM in \eqref{eq:OFDM_FIM} is invariant with respect to the parameter realization $\thetaVec$.
Therefore, equality holds in \eqref{eq:bound_comparison} and the capacity-MSE function $C(\Delta)$ coincides with the BCRB-based upper bound $C^{\mathrm{ub}}(\Delta)$.
Moreover, $C(\Delta)$ remains the same if one adopts a non-Bayesian framework with the standard CRB instead of the ECRB, as in \cite{hua2024}.
In that sense, the characterization in \eqref{eq:opt_ofdm} can be seen as a special case of \cite[eq.~(18)]{hua2024}, notwithstanding that the Bayesian framework in the present paper is more general.
\end{remark}
The communication-optimal trade-off point—arising when $\Delta$ in \eqref{eq:opt_ofdm} is sufficiently large so that the sensing constraint becomes inactive—is achieved by a water-filling subcarrier power allocation \cite[Sec.~9.6]{Cover}. 
On the other hand, the sensing optimal trade-off point is achieved with an equal power allocation \cite[Prop.~2]{hua2024}.
This follows from the fact that the MSE in \eqref{eq:OFDM_MSE} is convex in the power allocation vector. Specifically, by Jensen's inequality, we have
\begin{align}
    \sigma^2_{\bm{\upsilon}}\sum_{k=0}^{K-1} \frac{1}{\sigma^2_{\xRvVec,k}}  
    &\geq \sigma^2_{\bm{\upsilon}}  \frac{K}{\frac{1}{K}  \sum_{k=0}^{K-1}\sigma^2_{\xRvVec,k}} = \sigma^2_{\bm{\upsilon}} {K^2}/{P}
\end{align}
with equality when $\sigma^2_{\xRvVec,k}=P/K$ for all $k$.
More generally, the structure of the optimal solution beyond these two extremes has been studied extensively in \cite{hua2024}.
It is worth noting that the sensing constraint in \eqref{eq:opt_ofdm} requires non-zero power allocation across all subcarriers in order to yield a finite MSE $\Delta$.
This contrasts with the communication-optimal water-filling solution, which may deactivate some subcarriers. This phenomenon is further explored next.
\subsubsection*{Frequency-selective vs. frequency-flat communication channels}
\begin{figure}[t]
\centering
\subfloat[communication channels
\label{fig:comm_channels}]{
\hspace{-9mm}
\includegraphics[width=0.90\linewidth]{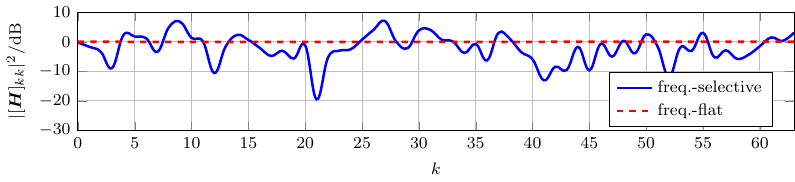}
} \\
\vspace{-4mm}
\subfloat[capacity--MSE trade-off \label{fig:ofdm_capacity_mse}]{
\includegraphics[width=0.90\linewidth]{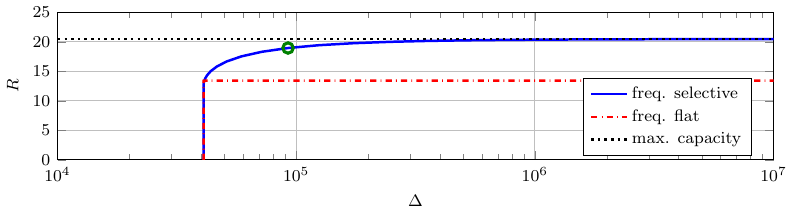}
} \\
\vspace{-4mm}
\subfloat[power allocation (we set $P = 64$) \label{fig:power_allocation}]{
\hspace{-4mm}
\includegraphics[width=0.90\linewidth]{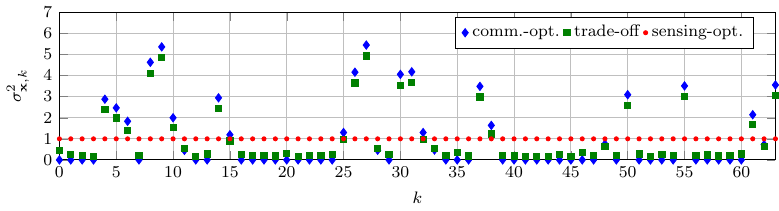}
}
\caption{Capacity–MSE trade-off in JCAS with OFDM.}
\label{fig:ofdm_result}
\end{figure}
Here, we numerically evaluate the capacity-MSE function. We set $K=64$ and study two cases: a frequency-selective communication channel and a frequency-flat communication channel. 
The former is meant to illustrate the trade-off between communication and sensing, while the latter will lead to no trade-off.
To this end, we set the entries of the time-domain channel impulse response $\bm{h}$ as
\begin{equation}
    h_k = \exp\left(-\frac{1}{2\alpha} k+\mathrm{i}\varphi_k\right), \ k = 0,1,\ldots,K-1
\end{equation}
where $\varphi_k$ is a uniformly distributed phase term. The channel model has an exponential power delay profile with delay spread $\alpha$. For the frequency-selective channel, we set $\alpha=10$. To get an almost frequency-flat channel, on the other hand, we chose $\alpha=0.1$. In both cases, we normalize the channel impulse response to have unit energy per subcarrier, i.e., $\lVert\bm{h}\rVert^2/K=1$. 
We define the communication and sensing SNR as $P/\sigma^2_{\bm{\omega}} = 10$ dB and $P/\sigma^2_{\bm{\upsilon}} = -10$ dB, respectively. 

The magnitude responses of the two channels are depicted in Fig.~\ref{fig:comm_channels}. 
The resulting capacity-MSE functions are illustrated in Fig.~\ref{fig:ofdm_capacity_mse}. We can see a clear trade-off between the capacity and the MSE in the frequency-selective channel case.
In the frequency-flat channel, on the other hand, there is no trade-off between 
communication and sensing, as one would expect.

In Fig.~\ref{fig:power_allocation}, we show the allocation of power among subcarriers for different points on the capacity-MSE curve.
In the frequency-selective scenario we consider, the communication\hyp optimal water-filling solution allocates zero power to subcarriers with insufficient SNR.
As a result, the communication-optimal point on the Pareto boundary is characterized by $\Delta \to \infty$, as shown in Fig.~\ref{fig:ofdm_capacity_mse}. 
For the intermediate trade-off point, indicated by a marker in Fig.~\ref{fig:ofdm_capacity_mse}, non-zero power is allocated to all subcarriers to guarantee a finite MSE, while still following the water-filling trend by allocating more power to subcarriers with higher SNR.
\section{Conclusion}
\label{sec:conc}
We considered the problem of joint communication and sensing in a basic Gaussian MIMO setting, where the sensing task is to estimate a random parameter vector. The random parameter vector, drawn according to a known prior distribution, is assumed to remain fixed throughout the transmission block.
We studied the fundamental trade-off between communication and sensing, formulated in terms of the rate of reliable communication (i.e., channel coding rate) against the MSE of parameter estimation.
The optimal rate-MSE trade-off in the asymptotic regime of large block-lengths, characterized in terms of a capacity-MSE function, was established, and a number of analytical properties were shown. 
Finally, case studies of exemplary scenarios illustrated this trade-off: one involving beamforming and DoA estimation, the other involving power allocation over OFDM subcarriers and tapped-delay channel response estimation.  

The problem we considered here can be extended in several directions. In the current paper, we focused our attention on non-adaptive (i.e., open-loop) transmission strategies. It is of interest to investigate the potential gains of adaptive (i.e., closed-loop) transmission strategies, where the next channel input is chosen depending on previous sensor observations.
Progress along these lines in the case of discrete parameters (i.e., hypothesis testing) has been reported in \cite{chang2023}, and it is of interest to extend some of these insights to problems involving continuous parameter estimation.
Another possible direction would be to extend the setting to multiple users, as well as multiple distinct targets, and study trade-off regions involving multiple rates and MSEs.
It is also of interest to refine the results and derive characterizations valid in the finite block-length regime, as has been done for channel coding problems \cite{Polyanskiy2010,Hayashi09,Yang2014}. As a first step in this direction, one can investigate refined asymptotics, such as second-order coding rates and tighter asymptotic MSE characterizations. Our achievability approach, based on Feinstein's lemma, provides a starting point for enabling such analysis.
Finally, relaxing the fixed-parameter assumption—without going as far as the i.i.d. varying regime—is also of interest. Ideally, one would like to investigate models where parameters evolve slowly, as in practical
scenarios. Initial attempts to model and study such scenarios in basic settings have been reported in \cite{Li2023capacity,Nikbakht2024,Lindstrom2025}.
\appendices
\section{}
\label{app:A}
\subsection{Proof of Lemma \ref{lemma:d_function}} \label{appendix:lemma_proof}
Here we show that the function $  d(\bm{Q}) \triangleq \trace{ \E \left[ \FIM(\thetaRvVec|\bm{Q})^{-1} \right] }$ is monotonically non-increasing and convex in $\bm{Q} \in \mathbb{S}_{+}^M$.
To this end, we first note that for any $\thetaVec$, the conditional FIM $\FIM(\thetaVec|\bm{Q})$ 
is positive semi-definite and linear in $\bm{Q}$, which can be verified directly from the definition, i.e., \eqref{eq:EFIM_Q_ij}.
Let $\bm{Q}_1$ and $\bm{Q}_2$ be two covariance matrices such that $\bm{Q}_1 \succeq \bm{Q}_2 $. 
We have 
\begin{align}
  \FIM(\thetaVec|\bm{Q}_1)  & = \FIM(\thetaVec|\bm{Q}_1 + \bm{Q}_2 - \bm{Q}_2) \\
  \label{eq:EFIM_app_1}
  & = \FIM(\thetaVec|\bm{Q}_2) +  \FIM(\thetaVec|\bm{Q}_1 - \bm{Q}_2)  \\
  \label{eq:EFIM_app_2}
  & \succeq  \FIM(\thetaVec|\bm{Q}_2)
\end{align}
where \eqref{eq:EFIM_app_1} is from linearity, while \eqref{eq:EFIM_app_2} holds since ${\bm{Q}_1 - \bm{Q}_2 \succeq \mathbf{0}}$ and hence ${\FIM(\thetaVec|\bm{Q}_1 - \bm{Q}_2)  \succeq \mathbf{0}}$.
Therefore, we have shown that $\FIM(\thetaVec|\bm{Q})$ is monotonically non-decreasing in $\bm{Q} \in \mathbb{S}_{+}^M$.
Note that \eqref{eq:EFIM_app_2} directly implies that ${\FIM(\thetaVec|\bm{Q}_2)^{-1} \succeq \FIM(\thetaVec|\bm{Q}_1)^{-1}}$, which in turn implies 
\begin{equation}
    d(\bm{Q}_2) \geq d(\bm{Q}_1).
\end{equation}
Therefore, we conclude that $ d(\bm{Q})$ is non-increasing.
For convexity, this holds since $\FIM(\thetaVec|\bm{Q})$ is linear in $\bm{Q}$, the inverse function $\FIM(\thetaVec|\bm{Q})^{-1}$ is convex, and then the expectation and trace are linear. Therefore, it can be easily verified that the overall function $d(\bm{Q})$ is convex. 
\subsection{Properties of \texorpdfstring{$C(\Delta)$}{C(Delta)}}
\label{appendix:subsec_properties_C}
We now show that the characterization of  $C(\Delta)$ in \eqref{eq:C_Delta_function} is continuous, non-decreasing, and concave in $\Delta$. 
First, it is clear from \eqref{eq:C_Delta_function} that the objective is non-decreasing in $\Delta$, since a solution under a smaller MSE constraint $\Delta$ is also feasible under a larger constraint.
To show concavity, let us consider two MSE constraint values $\Delta_1$ and $\Delta_2$, and let $\bm{Q}_1$  and $\bm{Q}_2$ be optimal solutions achieving $C(\Delta_1)$ and  $C(\Delta_2)$, respectively, in \eqref{eq:C_Delta_function}.
We have 
\begin{equation}
\label{eq:alpha_C_ub}
    \alpha C(\Delta_1) + (1-\alpha) C(\Delta_2) \leq
     \log \det \left( \bm{I} + \frac{1}{\sigma^2_{\bm{\omega}}}\bm{H} \big(  \alpha  \bm{Q}_1 + (1-\alpha) \bm{Q}_2) \bm{H}^{\herm} \right) 
\end{equation}
which holds due to the concavity of the Gaussian mutual information function. We observe that
\begin{align}
    d\big(\alpha  \bm{Q}_1 + (1-\alpha) \bm{Q}_2 \big) & \leq \alpha  d\big( \bm{Q}_1) + (1-\alpha) d ( \bm{Q}_2 ) \\
    \label{eq:d_alpha_ub}
    & \leq \alpha \Delta_1 + (1-\alpha) \Delta_2
\end{align}
where the first inequality is by convexity of $d(\bm{Q})$, while the second inequality follows from the optimality (and hence feasibility) of $\bm{Q}_1$  and $\bm{Q}_2$ for their corresponding problems. 
The inequality in \eqref{eq:d_alpha_ub} implies that $\alpha  \bm{Q}_1 + (1-\alpha) \bm{Q}_2 $ is a feasible solution for \eqref{eq:C_Delta_function} with an MSE constraint $\Delta = \alpha \Delta_1 + (1-\alpha) \Delta_2$.
Therefore, \eqref{eq:alpha_C_ub} is further upper bounded by $C\big(  \alpha \Delta_1 + (1-\alpha) \Delta_2 \big)$, which proves concavity.
Continuity, on interior points of the domain, follows from concavity.

\section{Proof of Lemma~\ref{lemma:covariance_concentration}}
\label{app:B}
Starting with the semi-positive definite constraint on the sequence $\tilde{\xRvVec}^N \in \bm{\mathcal{B}}_{N,\delta}(\bm{Q})$, i.e.,
\begin{equation}
    \bm{Q} - \delta \bm{I} \preceq \hat{\bm{Q}}(\tilde{\xRvVec}^N) \preceq \bm{Q} + \delta \bm{I}
\end{equation}
the following scalar inequality holds by the definition of positive semi-definiteness
\begin{equation*}
    \bm{a}^\herm\left(\bm{Q}-\delta\bm{I}\right)\bm{a} \leq   \bm{a}^\herm \left(\frac{1}{N}\sum_{i =1}^N \tilde{\xRvVec}_i \tilde{\xRvVec}_i^{\herm}\right)\bm{a} \leq \bm{a}^\herm \left(\bm{Q}+\delta\bm{I}\right)\bm{a}.
\end{equation*}
This is equivalently written as
\begin{equation}
    \bm{a}^\herm\bm{Q}\bm{a} -\delta\lVert\bm{a}\rVert^2 \leq   \frac{1}{N}\sum_{i=1}^N |\bm{a}^\herm\tilde{\xRvVec}_i|^2 \leq  \bm{a}^\herm\bm{Q}\bm{a} +\delta\lVert\bm{a}\rVert^2
    \label{eq:covariance_inequality}
\end{equation}
for any $\bm{a} \in \C^M$ and some $\delta > 0$.
Since $\bm{Q} \in \mathbb{S}_+^M$ is Hermitian positive semi-definite, there exists some matrix $\bm{D} \in \C^{r \times M}$ 
with $r \triangleq \rank(\bm{Q}) \leq M$ such that $\bm{Q} = \bm{D}^\herm \bm{D}$.

Let $\mathbf{u}_i \sim \CN(\bm{0},\bm{I})$ and let $\tilde{\xRvVec}_i \triangleq \bm{D}^{\herm}  \mathbf{u}_i$.
For $r < M$, i.e.,  $\bm{Q}$ is  rank-deficient, the nullspace of $\bm{D}$ contains non-zero vectors.
When $\bm{a}$ is in the nullspace of $\bm{D}$, i.e., $\bm{D} \bm{a}=\bm{0}$, \eqref{eq:covariance_inequality} becomes
\begin{equation}
-\delta\lVert\bm{a}\rVert^2 \leq  0 \leq \delta\lVert\bm{a}\rVert^2
\end{equation}
which trivially holds true.
When $\bm{Q}$ is full-rank, $\bm{D}$ is a full-rank square matrix, and the null space consists of only the zero vector.
When $\bm{a} \neq \bm{0}$ is not in the nullspace of $\bm{D}$, i.e., $\bm{D} \bm{a} \neq \bm{0}$, we can divide \eqref{eq:covariance_inequality} by $\bm{a}^\herm\bm{Q}\bm{a}$, which yields
\begin{equation}
    \label{eq:interm_form}
     1 -\frac{\delta\lVert\bm{a}\rVert^2}{\bm{a}^\herm\bm{Q}\bm{a}} \leq \frac{1}{N}  \sum_{i=1}^N \left|\frac{\bm{a}^\herm\tilde{\xRvVec}_i}{\sqrt{\bm{a}^\herm\bm{Q}\bm{a}}} \right|^2 \leq 1 +\frac{\delta\lVert\bm{a}\rVert^2}{\bm{a}^\herm\bm{Q}\bm{a}}.
\end{equation}
Since \eqref{eq:covariance_inequality} must be satisfied for any $\bm{a} \in \C^M$, a sufficient condition is obtained using the bound
\begin{equation}
    \bm{a}^\herm \bm{Q} \bm{a} \leq \lambda_{\max}(\bm{Q}) \lVert \bm{a} \rVert^2
\end{equation}
where $\lambda_{\max}(\bm{Q})$ is the largest eigenvalue of $\bm{Q}$.
Defining $\upalpha_i \triangleq (\bm{a}^\herm\bm{Q}\bm{a})^{-\frac{1}{2}} \bm{a}^\herm \tilde{\xRvVec}_i$, we get
\begin{equation}
    \label{eq:interm_form2}
    1-\frac{\delta}{\lambda_{\max}(\bm{Q})} \leq \frac{1}{N}\sum_{i=1}^{N} \left|\upalpha_i\right|^2\leq 1+\frac{\delta}{\lambda_{\max}(\bm{Q})}.
\end{equation}
We observe that $\upalpha_i \sim \CN(0,1)$. Setting $\delta^{\prime} \triangleq \delta / \lambda_{\max}(\bm{Q}) > 0$, we can rewrite \eqref{eq:interm_form2} as
\begin{equation}
    \left\lvert \frac{1}{N} \sum_{i=1}^N \lvert \upalpha_i \rvert^2 - 1 \right\rvert \leq \delta^{\prime}
\end{equation}
where $|\upalpha_i|^2 \sim \mathrm{Exp}(1)$ has an exponential distribution. Let $\mathrm{S}_N \triangleq \frac{1}{N} \sum_{i=1}^{N} |\upalpha_i|^2$ denote the empirical mean of $N$ i.i.d. exponential random variables.
The mean and the variance of $\mathrm{S}_N$ are given by $\E[\mathrm{S}_N] = 1$ and $\E[(\mathrm{S}_N - 1)^2] = 1/N$, respectively.
Finally, for any $\delta > 0$
\begin{align}
    \Prb[\tilde{\xRvVec}^N \notin \bm{\mathcal{B}}_{N,\delta}(\bm{Q})]
    &\leq \Prb[|\mathrm{S}_N - 1| \geq \delta^{\prime}] \label{eq:upperbound} \\
    &\leq \frac{1}{\delta^{\prime 2} N} \label{eq:chebyshev}
\end{align}
using the sufficient condition in \eqref{eq:interm_form2} as an upper bound in \eqref{eq:upperbound} and Chebyshev's inequality in \eqref{eq:chebyshev}. As $N \to \infty$, we get $\Prb[\tilde{\xRvVec}^N \in \bm{\mathcal{B}}_{N,\delta}(\bm{Q})] \to 1$, which completes the proof.

\section{BCRB-Based Capacity-MSE Bound}
\label{app:C}
In this appendix, we show how to obtain the capacity-MSE upper bound in \eqref{eq:bcrb_opt} using the BCRB lower bound \cite{vanTrees2013}. 
The derivation builds on the converse proof in Section \ref{subsec:converse}.

To define the BCRB, we first introduce the Bayesian information matrix associated with estimating $\thetaRvVec$ from $\zRvVec^N$, given that $\xRvVec^N = \xVec^N$ is transmitted. This is written as
    \begin{equation}
        \label{eq:BIM}
        \FIM_{\mathrm{B}}(\xVec^N) \triangleq \E[\FIM(\thetaRvVec|\xVec^N)] +  \FIM_{\mathrm{P}}
    \end{equation}
    where $\FIM_{\mathrm{P}} \triangleq \E \left[ [ \nabla_{\thetaVec} \ln f_{\thetaRvVec}(\thetaRvVec) ] \left[ \nabla_{\thetaVec} \ln f_{\thetaRvVec}(\thetaRvVec) \right]^{\trans} \right]$ is the prior information matrix.
    The BCRB inequality of van Trees \cite[Sec.~4.3]{vanTrees2013} states the MSE is lower bounded as
    \begin{equation}
         N \trace{\mathbf{\Sigma} \left(\xVec^N , \hat{\thetaVec}_N^{\star} \right) } \geq  \trace{ \left[ \frac{1}{N}\FIM_{\mathrm{B}}(\xVec^N) \right]^{-1} } \trace{ \left[ \E \left[ \bm{J} \big( \thetaRvVec | \hat{\bm{Q}}(\xVec^N) \big) \right] + \frac{1}{N} \FIM_{\mathrm{P}} \right]^{-1} }.
    \end{equation}
    Unlike the lower bound in \eqref{eq:ECRB_Q}, this bound holds not only asymptotically but for every $N$.
    Adopting this in the converse proof in Section \ref{subsec:converse}, and continuing from \eqref{eq:MSE_LB_1}, we have 
    \begin{align}
    \label{eq:BCRB_LB_1}
    \Delta  + \epsilon 
    & \geq \frac{1}{M_N}  \sum_{w \in \mathcal{M}_N} \trace{  \left[ \E  \left[ \bm{J} \big( \thetaRvVec | \hat{\bm{Q}}(\xVec^N(w)) \big) \right] + \frac{1}{N} \FIM_{\mathrm{P}} \right]^{-1} } \\
    \label{eq:BCRB_LB_2}
    & \geq \trace{  \left[ \E\left[ \bm{J}(  \thetaRvVec | \bar{\bm{Q}} ) \right] +  \frac{1}{N} \FIM_{\mathrm{P}} \right]^{-1} }  \\
    & \geq \trace{  \E\left[ \bm{J}(  \thetaRvVec | \bar{\bm{Q}}    ) \right]^{-1} }    -  \epsilon
    \label{eq:BCRB_LB_3}
\end{align}
where \eqref{eq:BCRB_LB_2} holds by convexity and Jensen's inequality, and \eqref{eq:BCRB_LB_3} holds for sufficiently large $N$ by continuity and the fact that $\FIM_{\mathrm{P}}$ is constant, and hence $ \frac{1}{N} \FIM_{\mathrm{P}} $ vanishes.

It is also possible to adopt the Miller--Chang version of the BCRB \cite{Miller1978}, as done in \cite{xiong2023}. In this formulation, the random codeword  $\xRvVec^N$ is treated as a nuisance parameter, and the average BCRB with respect to $\xRvVec^N$  is considered. Adapting this to our setting amounts to replacing the maximal MSE risk (over codewords) $\xi_N$  with its average counterpart  $\xi_N^{\mathrm{av}}$ (see Section \ref{subsubsec:estimation}). 
Nevertheless, since $\xRvVec^N$ is uniformly distributed on the codebook as the message is uniform, the previous result still applies. In particular, \eqref{eq:BCRB_LB_1} onward remain valid.

\balance
\bibliographystyle{IEEEtran}
\bibliography{bibliography}
\end{document}